\PassOptionsToPackage{dvipsnames}{xcolor}
\RequirePackage{xcolor}
\documentclass[openacc]{rsproca_new}%%%%where rsproca is the template name

\usepackage[colorinlistoftodos,textsize=footnotesize]{todonotes}

\newcommand{\pder}[2][]{\frac{\partial#1}{\partial#2}}
\usepackage{subfigure}
\usepackage{lmodern}
\usepackage{siunitx}
\usepackage{booktabs}
\usepackage{etoolbox}
\usepackage{comment}

\titlehead{Research}

\begin{document}

%%%% Article title to be placed here
\title{Using surface acoustic waves to drive thin film flow over an obstacle}

\author{%%%% Author details
Y. Li$^{1}$\thanks{First and second authors contributed equally, with Li focusing on experiments and Fasano on modeling and simulations.}, M. Fasano$^{2}$\footnotemark[1], A. R. Einhorn$^{1}$, J. A. Diez$^{3}$, O. Manor$^{1}$, L. J. Cummings$^{2}$, and L. Kondic$^{2}$}

%%%%%%%%% Insert author address here
\address{$^{1}$ Department of Chemical Engineering, Technion - Israel Institute of Technology, Haifa 32000, Israel\\
$^{2}$ Department of Mathematical Sciences and Center for Applied Mathematics and Statistics, New Jersey Institute of Technology, Newark, New Jersey 07102, USA\\
$^{3}$Instituto de F\'{\i}sica Arroyo Seco, Universidad Nacional del Centro de la Provincia de Buenos Aires and CIFICEN-CONICET-CICPBA, 7000 Tandil, Argentina}

%%%% Subject entries to be placed here %%%%
\subject{Acoustics, Computational physics, Mathematical modelling}

%%%% Keyword entries to be placed here %%%%

%%%% Insert corresponding author and its email address}
\corres{Insert corresponding author name\\
\email{manoro@technion.ac.il}}

%%%% Abstract text to be placed here %%%%%%%%%%%%
 \begin{fmtext} 
\begin{abstract}
We study a new paradigm for ultrasonic driven object coating by using a model system where MHz-level surface acoustic waves (SAWs) drive the spreading of a silicone oil film atop topographical obstacles. We use experiments to show that nanometer-amplitude SAWs, propagating in the substrate of a piezoelectric actuator, propel macroscopic oil films to climb and traverse solid obstacles placed on the actuator. The oil dynamics reveal rich coupling between ultrasonic forcing, capillarity, and gravity; the balance of which determines coating success. We formulate a simplified two-dimensional theoretical model that incorporates obstacle geometry directly in the oil thin-film evolution equation, introducing a new representation of acoustic streaming in the presence of substrate height variations. Despite the simplifications inherent in the modeling, simulations show qualitative agreement with the experiments, providing evidence that the model captures the key physics.
\end{abstract}

%%%%%%%%%%%%%%%%%%%%%%%%%%%
 
\rsbreak

%%%%%%%%%% Insert the texts which can accomdate on firstpage in the tag "fmtext" %%%%%

\section{Introduction}
    Significant interest has emerged in coating flows over topographical features due to applications in many industrial processes. Such flows are employed in technological systems for the actuation of microfluidic platforms~\cite{Atencia:2005bu,Stone:2004kg,Whitesides:2006jj} and the cooling of electronic circuits~\cite{Amon:2001ji,BarCohen:2006if}; they also occur in desalination of saltwater~\cite{Fletcher:1974jz}, and the coating behavior of liquid paints, among other areas. Often flows are driven by gravity~\cite{gaskell04gravity} and/or surface tension~\cite{stillwagon1988fundamentals,Howell2003,Schwartz1995}, situations where the governing equations are well-known and have been described in comprehensive reviews such as those by Oron et al.~\cite{oron_rmp97} or Craster and Matar \cite{cm_rmp09}. 
    
    However, other physical mechanisms may drive coating flows. It has been shown that mechanical vibrations in a solid substrate, most commonly MHz-frequency Rayleigh surface acoustic waves (SAWs), can produce coating flows of thin films atop the solid via interfacial phenomena: acoustic radiation pressure and convective contributions in SAW-induced boundary layer flow~\cite{Shiokawa:1989tg,morozov_vibration-driven_2018,altshuler2015spreading,Horesh}. The former mechanism is an excess quasi-steady stress at the free surface of a film that acts over timescales longer than the SAW period. It results from the advection of momentum of ultrasound that diffracts (leaks) off the SAW and deflects off the free film surface~\cite{King:1934tp, Chu, Hasegawa:2000vy}. The latter is a quasi-steady drift flow, generated at long times in a boundary layer flow, also known as Rayleigh streaming or the Rayleigh law of streaming, due to the convection of momentum near the solid surface that supports the SAW~\cite{LordRayleigh1884,LIGHTHILL:1978p12, longuet-higgins_mass_1953, manor_yeo_friend_2012, yeoannurev, morozov_extended_2017}. The thickness of the boundary layer flow, i.e., the viscous penetration of the SAW into the liquid, $\delta=\sqrt{2\mu/(\rho\omega)}$, is usually sub-micron for ordinary liquids in the MHz SAW regime, where $\mu$, $\rho$, and $\omega$ denote the liquid viscosity, the liquid density, and the angular frequency of the SAW, respectively. At locations where the liquid thickness is comparable to or larger than the wavelength of the ultrasound that diffracts off the SAW, the dominant acoustic streaming mechanism is known as Eckhart streaming. Eckhart streaming signifies a drift of mass, resulting from variations in the intensity of the ultrasound that diffracts (leaks) off the SAW in the fluid~\cite{Eckart:1948to,Nyborg:2004p312}.
    
    Several studies have examined bulk (Eckart) acoustic streaming in water films and its contribution to their displacement~\cite{Brunet07,Brunet09,yeoannurev}. Moreover, a self-consistent theory was developed recently for the dynamics of thick free surface films due to bulk streaming and was verified by experiments that employed millimeter-thick silicone oil films atop the flat substrate of 20~MHz frequency SAW actuators~\cite{Li2024}. In addition to the acoustic stress in the film, the analysis accounted for the capillary and gravitational stresses that shape the liquid/air interface and dictate its dynamics. 

    Here, we study SAW-induced coating flows of thick liquid films over obstacles as a paradigm for using ultrasound for coating processes. We focus on experiments and a long-wave theory to investigate the free-surface evolution of thick films that support Eckhart streaming over topographical features via a SAW propagating in an underlying solid substrate. To facilitate such a development, we build upon the model developed by Fasano et al. \cite{Li2024} to incorporate the additional contribution to the gravitational and curvature stresses of a topographical feature, taking an approach consistent with the classical work of Stillwagon et al.~\cite{stillwagon1988fundamentals} and the more recent work of Park and Kumar\cite{park2017droplet}. 
    
    The remainder of the paper is structured as follows: In Sec. \ref{sec:experiment} we discuss physical experiments that demonstrate the phenomenon of thick silicone oil films driven over various obstacles under the action of propagating, 20~MHz frequency, Rayleigh-type SAWs. In Sec. \ref{sec:model} we present a corresponding mathematical model,  and in Sec. \ref{sec:numerical_results} we show its application to the experimental setup. We summarize the model and initial conditions in Sec. \ref{sec:numEqs}, before comparing our theory to the two obstacle scenarios in our experiments: a ramp obstacle in Sec. \ref{sec:ramp} and a bump obstacle in Sec. \ref{sec:bump}. In Sec. \ref{sec:discussion} we compare our experimental and theoretical results, and in Sec. \ref{sec:conclusions} we summarize our findings and draw conclusions. Technical details and background are relegated to appendices: Appendix \ref{app:SpreadingModel} summarizes the theoretical model without an obstacle present that was developed by Fasano et al.~\cite{Li2024} and forms a basis for the extended model presented here, and Appendix \ref{app:lub} describes the long-wave approximation used to derive the model in which topographical obstacles are present. Additionally, videos of selected experiments and animations of representative computational results are available upon request.
    
\section{Experiment} \label{sec:experiment}

\subsection{Experimental system}
\begin{figure}[htp]
    \centering
    \includegraphics[width=0.8\columnwidth]{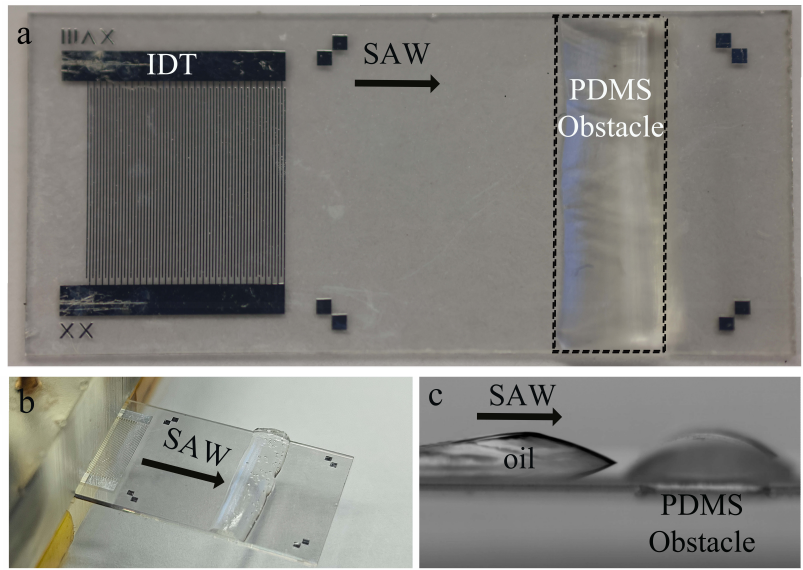}
    \caption{ (a) Top view of the SAW actuator used in the experiment (dimensions: 24.5 mm × 10.8 mm × 0.5 mm), comprising an inter-digital transducer to the left made from metal electrodes fabricated atop the piezoelectric lithium-niobate substrate of the actuator, and a PDMS obstacle placed atop the SAW actuator (to the right). (b) The SAW actuator, supporting a 3D printed PDMS obstacle viewed from above, attached to a power source (to the left) and (c) a side view of the same SAW actuator supporting the PDMS obstacle (to the right) and an oil film that spreads over the SAW actuator and toward the obstacle under the action (and in the direction) of the SAW in the solid substrate of the actuator. }
    \label{fig:system}
\end{figure}
Figure \ref{fig:system} shows the experiment, where an 8 $\mu$l oil film is driven by a 20~MHz frequency surface acoustic wave (SAW) toward a PDMS obstacle. The nanometer-amplitude mechanical wave is generated by the SAW actuator, comprised of a 5 nm titanium / 1 $\mu$m aluminum interdigitated transducer (IDT; from which the SAW emanates) fabricated atop lithium niobate (LiNbO$_3$, 128° Y-cut, X-(SAW) propagating, Roditi International, UK) by standard photolithography (Fig. \ref{fig:system}(d)). The actuator is powered using pogo pins (BC201403AD, Interconnect Devices, Inc.) assembled in a 3D-printed elastomeric stage (Fig. \ref{fig:system}(a)), which holds the actuator and is connected to a signal generator (R\&S SMB100A microwave signal generator) and an amplifier (model A10160, Tabor Electronics Ltd.). 

The PDMS obstacle, shown on the right of Figure \ref{fig:system}(b-c) is made of silicone-based elastomer (SYLGARD™ 184 Silicone Elastomer Kit, Dow, Australia) with a mixture ratio of 8 (pre-polymer): 1 (curing agent). It is attached to the lithium niobate substrate, about $5$~mm away from the IDT, by pre-treating the substrate for 10 minutes using an air plasma cleaner (PDC-002 Plasma cleaner, Harrik Plasma, New York, U.S.). We consider two obstacle geometries: (i) cylindrical cap `bump' obstacles, of several heights (as shown in Figure~\ref{fig:system}) and (ii) a prism-shaped `ramp' obstacle, of slope approximately $42^\circ$ relative to the flat substrate of height $3.2$ mm and width $3.5$ mm.

We employ a top-view camera (Dino-Lite Digital Microscope Premier) and a higher-precision camera (Data Physics; OCA 15Pro, CMOS id: UI-3360CP-M-GL) for side-view documentation of the obstacle coating. The side view camera provides a clear contrast between the silicone oil and the PDMS obstacle enabling us to track the location of the oil film front on the obstacle, although the high resolution of the camera comes at the expense of a limited field of view. The top-view camera allows us to track the overall coating process and determine how long it takes for the oil to overcome the obstacles.  

In each experiment, a sessile film of silicone oil ($\nu=50$ cSt, 378356, Sigma-Aldrich) is deposited on the surface of the substrate using a pipette, approximately $2$ mm from the obstacle. The SAW is then activated by applying a sinusoidal voltage signal of $f=20$ MHz frequency. The applied voltage is proportional to the measured normal displacement amplitude, $A_n$, of the SAW at the surface of the actuator, which is in contact with the oil film. Once power is applied to the actuator, the oil film begins spreading along the path of the SAW. Experiments were performed by placing an oil volume of $V_d=8~\mu$l or $16~\mu$l; in the latter case, this was done by placing an oil volume of 8 $\mu$l atop the actuator twice, as explained below.  

\subsection{Ramp obstacle}

A typical ramp obstacle experiment is shown in Fig.~\ref{fig:ramp}: the ramp is a triangular prism with a right angle, see Fig.~\ref{fig:ramp}(a). The horizontal width of the ramp (in the direction of SAW propagation) is 3.5 mm, the height is 3.2 mm and the angle of the slope with respect to the substrate is $42^\circ$. In the experiment, we placed 8 $\mu l$ of silicone oil on the actuator and induced a SAW of varying strength (as shown in Fig.~\ref{fig:ramp}(b)). Under the influence of SAW, the silicone oil moves in the direction of the propagating wave and climbs up the ramp (Figs.~\ref{fig:ramp}(b-f)). The PDMS elastomer has a similar chemistry and density to the silicone oil and thus a similar acoustic impedance. Hence, the ultrasonic wave that diffracts off the SAW in the actuator propagates also through the PDMS obstacle, driving coating motion of the silicone oil film over it. The ultrasonic energy available for climbing will decrease as the silicone oil ascends higher on the ramp due to the dissipation of this energy in the PDMS structure. Eventually, the oil either reaches the top of the ramp (Fig.~\ref{fig:ramp}(f)) or reaches a steady-state height because the reduced ultrasonic energy is insufficient to support further climbing. (Complete experimental videos available upon request.)

\begin{figure}[ht]
    \centering
    \includegraphics[width=0.8\linewidth]{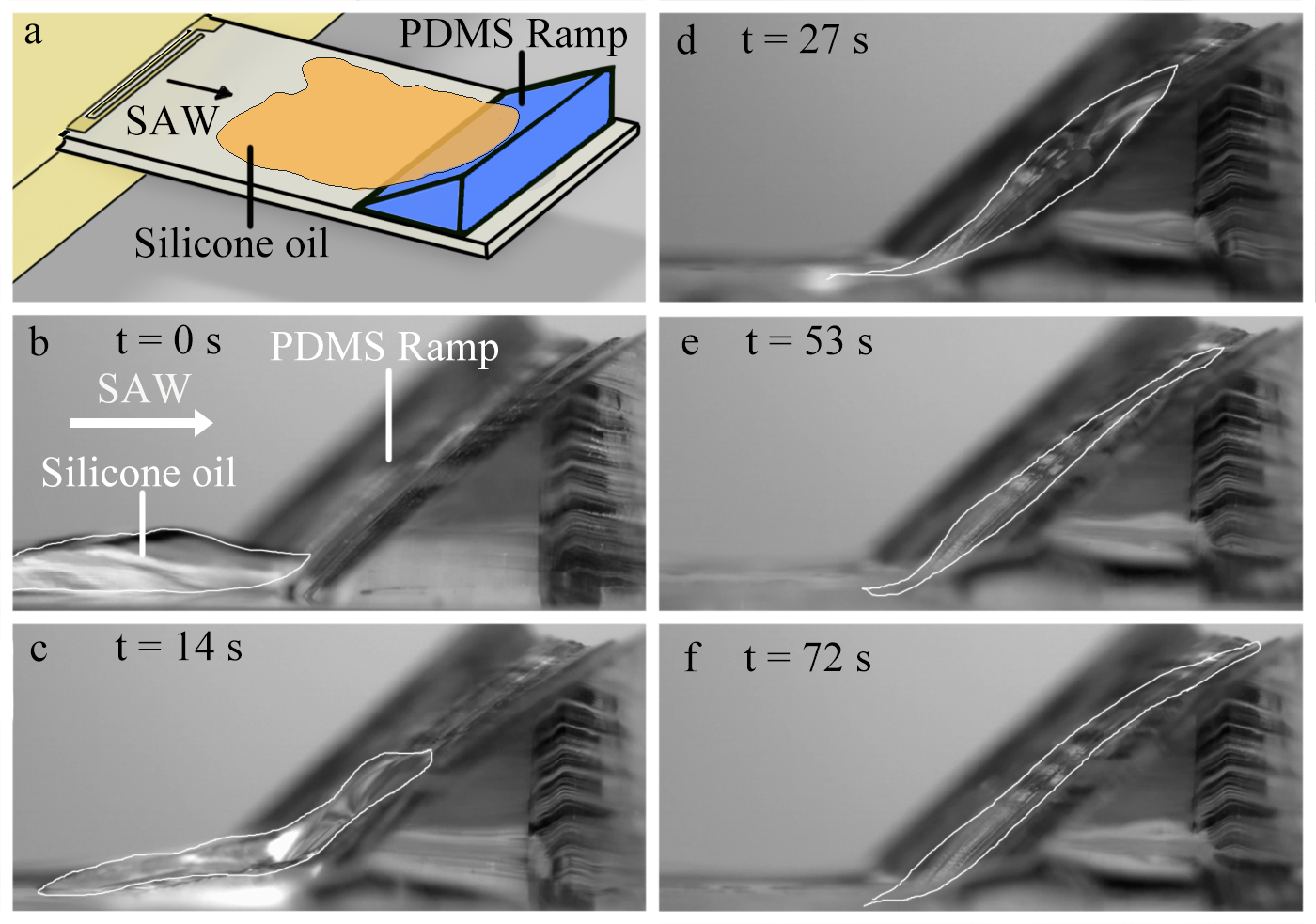}
    \caption{Side view snapshots of a typical experiment ($A_n$=1.43 nm), where oil climbs over a ramp-shaped obstacle, indicated by the sketch in (a). During the experiment, (b) initially, the oil film (highlighted using a thin, white curve) moves along the upper surface of the SAW transducer to make contact with the obstacle; the oil film then deforms and (c,d,e) climbs up the obstacle and (f) reaches the peak. Time $t$ is measured from the moment the oil comes in contact with the ramp. (Full experimental videos available upon request.) }
    \label{fig:ramp}
\end{figure}

To better understand the climbing process, an analysis of the dynamic moving oil front is necessary. By carefully analyzing the experimental side-view videos (using ImageJ), we find the climbing height of the oil film on the ramp under different experimental conditions. Figure~\ref{fig:exp_ramp_over_time} displays such height versus time (measured from the time when the silicone oil front first touches the rear of the ramp) for several different SAW amplitudes; part (a) shows selected results for the whole range of SAW amplitudes used ($A_n$ in the range 0.52~nm--1.69~nm); while part (b) presents the same results with log-scaled time.  
%Except for the case of $A_n=0.52$~nm, the silicone oil film successfully surpasses the top of the ramp, reaching a final height of 3.2~mm. 
The total climbing time varies with SAW amplitude $A_n$; larger $A_n$ values result in shorter climbing times as expected, since $A_n$ is proportional to the square root of the ultrasonic power supplied to the oil film. 
%{\color{red} ** Need to clarify whether results in Fig. 3 are for 8mm$^3$ oil only, or 8+8. As written it strongly suggests the former, but Mark thought it was the latter. Needs rewriting if so, and moving to after the paragraph below that explains the 8+8 protocol. **}
\begin{figure}[htp]
    \centering
    \subfigure[]{
\includegraphics[width=0.42\linewidth]{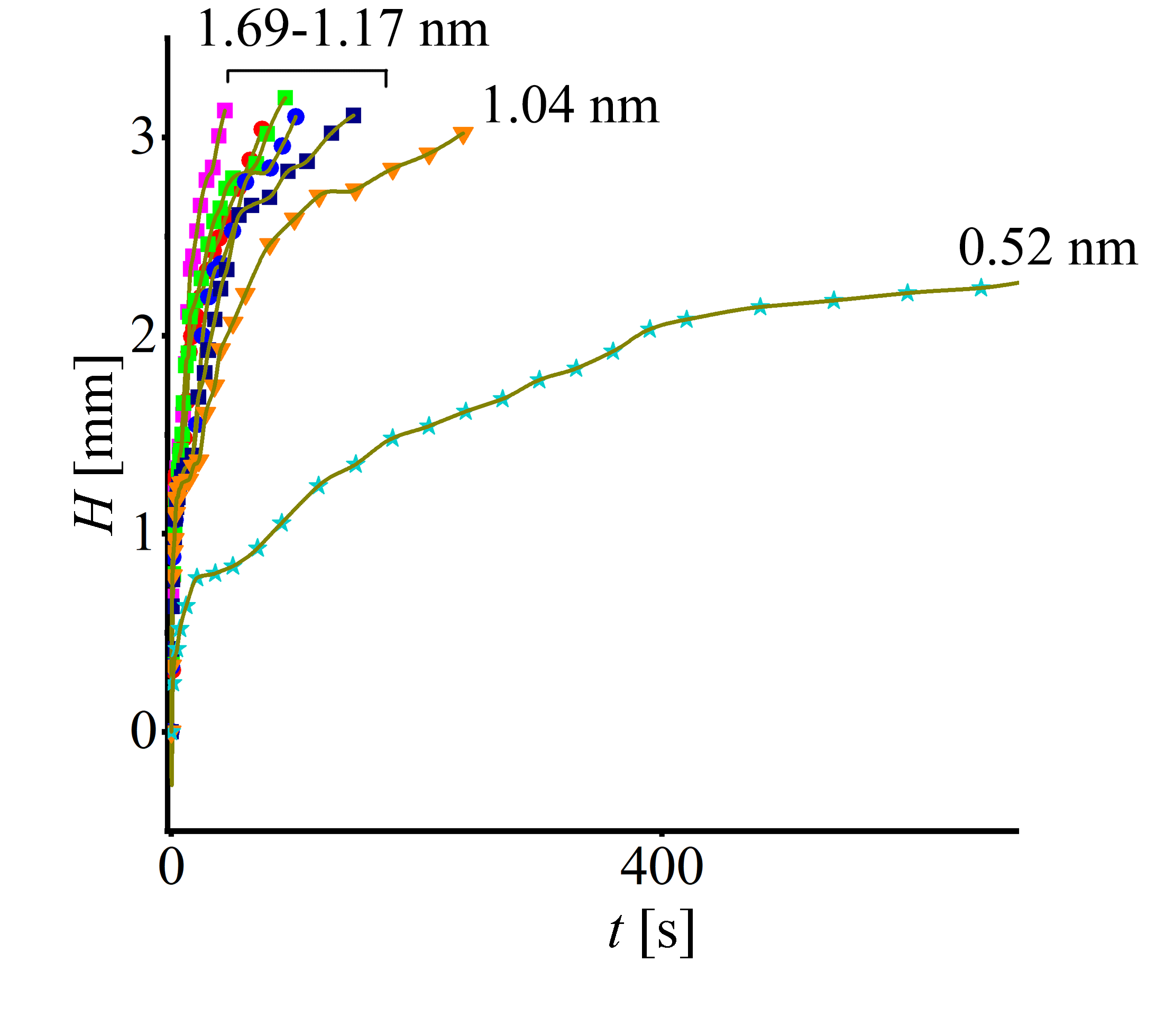}}
 \subfigure[]{
\includegraphics[width=0.42\linewidth]{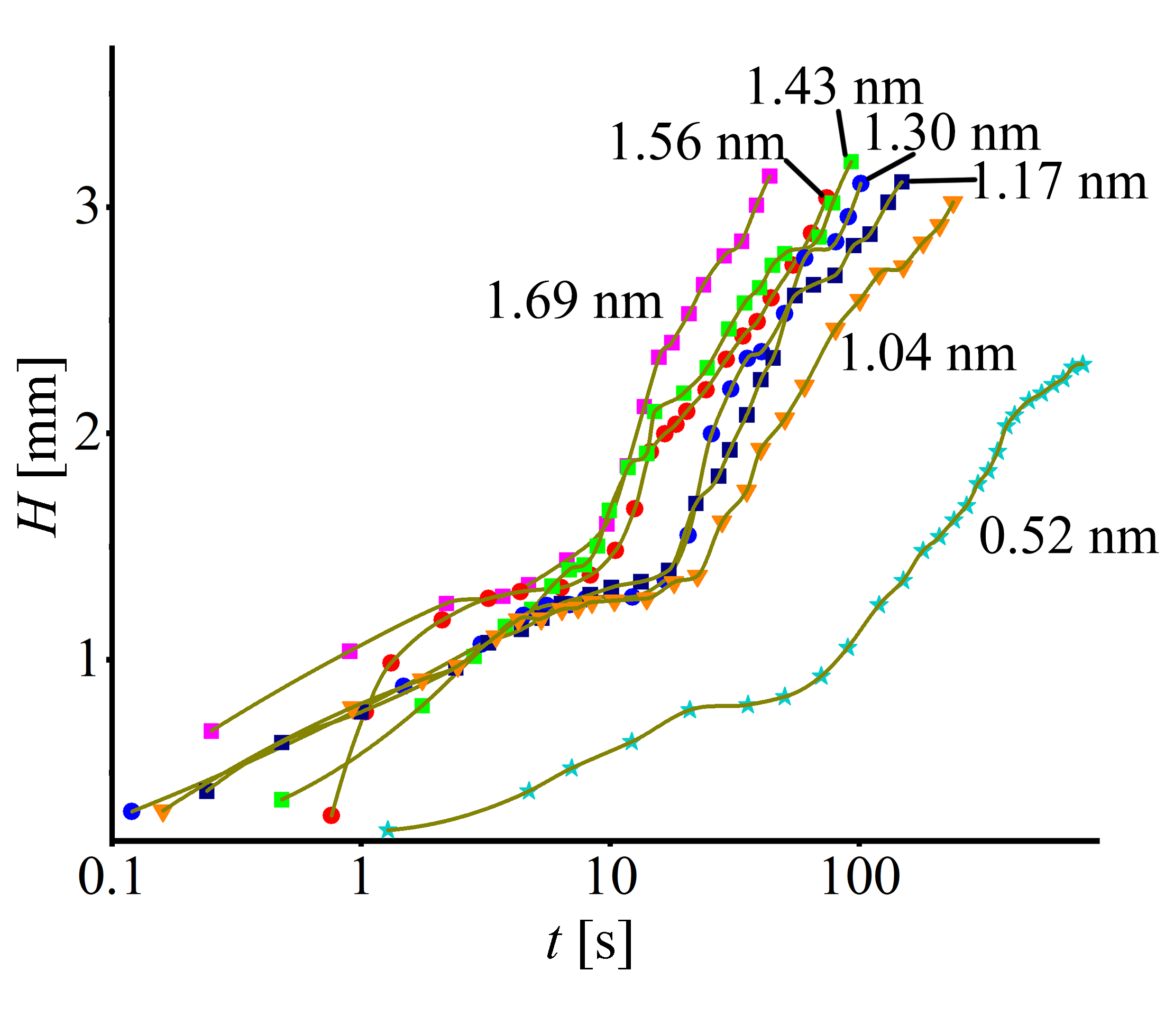}}
    \caption{Time variation of the climbing height of the film atop the ramp for (a) $A_n$ = 0.52 nm to 1.69 nm and (b) the same results with a logarithmic time axis, indicating a change in the mechanism that drives the climb above a height of approximately 1 mm. Symbols are experimental data and the connecting curves are guides for the eye. The uncertainty in climb height is 5\% based on the spatial resolution of the camera.    }   
    \label{fig:exp_ramp_over_time}
\end{figure}

The tendency of the oil to flow around rather than over the ramp (which cannot occur in the 2D simulations that will be described in Sec.~\ref{sec:numerical_results}) results in a diminished volume of oil at the base of the ramp. This, in turn, leads to a decreased capability of the oil film to climb, possibly indicating that power transfer to the oil film from the SAW is more efficient when the oil is in direct contact with the SAW actuator.  To ensure that an adequate volume of oil remains in contact with the actuator, we conducted another set of experiments in which, once a steady-state climbing height was reached, an additional 8 $\mu$l of silicone oil was added at the bottom of the ramp, encouraging further climbing. Figure~\ref{fig:exp_height} shows the maximum climbing height, $H_\text{max}$, both with (red dots) and without (blue dots) additional silicone oil, for all experiments conducted. The results of Fig.~\ref{fig:exp_ramp_over_time} represent a subset of the results shown here; they are the experiments for which the largest climbing height was attained for the respective $A_n$-values (highest blue dots in Fig.~\ref{fig:exp_height}). We observe that the additional oil enhances climbing (the red dots are associated with larger $H_{\rm max}$-values), so that, for $A_n\gtrsim~1$ nm, the oil film reaches the top of the obstacle.

\begin{figure}[htp]
    \centering
    \includegraphics[width=0.54\columnwidth]{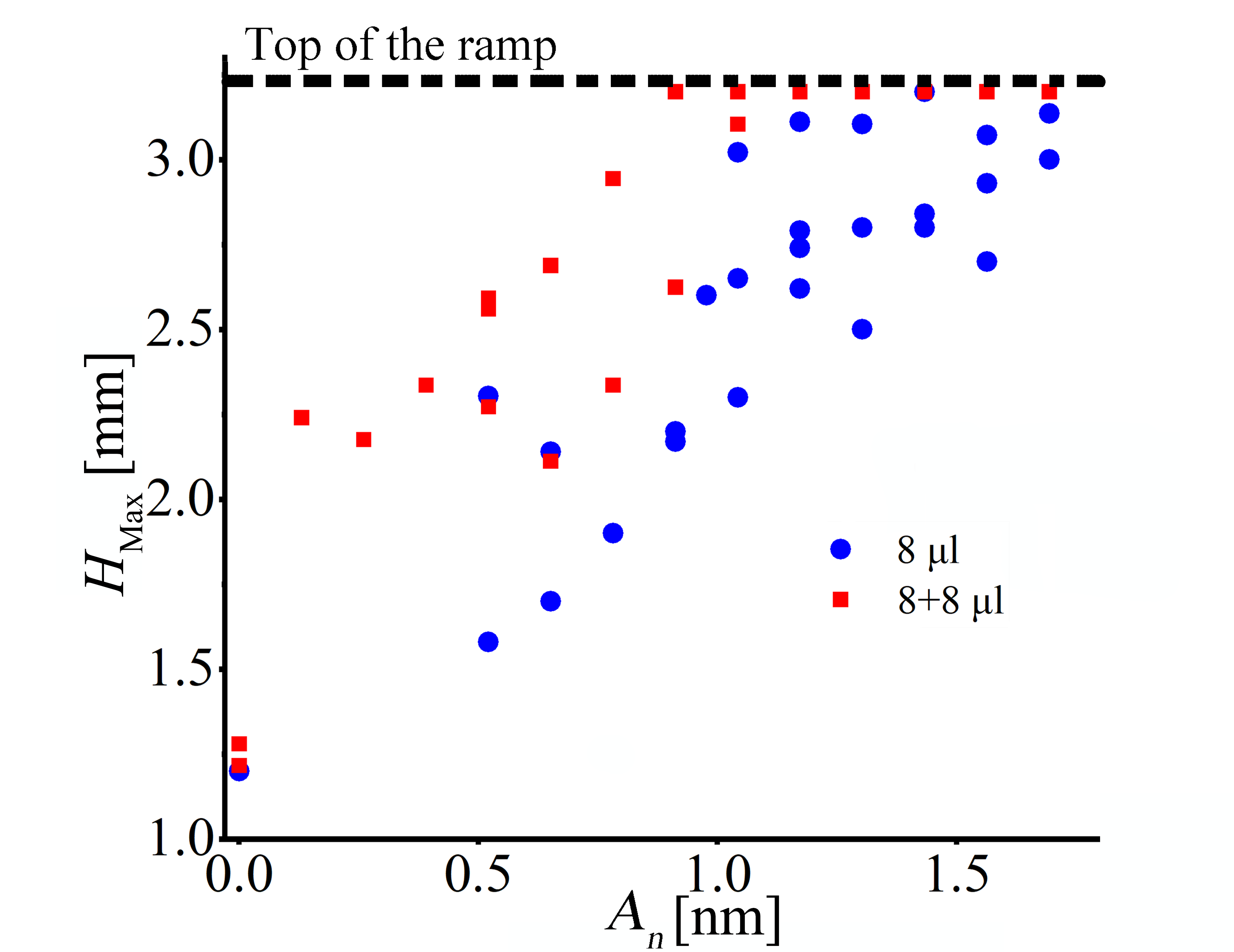}
    \caption{Maximum climbing height, $H_\text{max}$, of the film on the ramp, vs SAW normal amplitude displacement ($A_n$), for 8 $\mu$l of silicone oil volume (blue dots) and for 8+8 $\mu$l of oil volume (red squares). The dashed line shows the top of the ramp. The uncertainty in climb height is 5\% based on the spatial resolution of the camera.  }
    \label{fig:exp_height}
\end{figure}

\subsection{Bump obstacle}
To better understand how substrate topography affects spreading, we also consider a different obstacle shape: a bump.  These obstacles are cylindrical cap structures, whose height, $h_{\rm o}$, varies from $0.26$~mm to $1.1$~mm. The bump width, $w_{\rm o}$ (distance from the rear to the front of the bump in the direction of SAW propagation and the motion of the oil film) is in the range of $2~-~3$ mm, with variability caused by the PDMS deposition process. We note that both $h_{\rm o}$ and $w_{\rm o}$ may vary along and across the bump, leading to variability of the bump properties and, therefore, of the reported results.
 
We apply SAW to propel the silicone oil toward and over the bump. The process is captured using two cameras: the side view captures the dynamics (shown in Fig. \ref{fig:exp_bump}(b1-e1)), while the top view allows us to determine the overall time needed to traverse the bump, defined as the time elapsed between the oil reaching the near side (the rear; Fig.~\ref{fig:exp_bump}(b2)) and the far side (the front; Fig.~\ref{fig:exp_bump}(e2)) of the bump. 

\begin{figure}[htp]
    \centering
    \includegraphics[width=0.65\columnwidth]{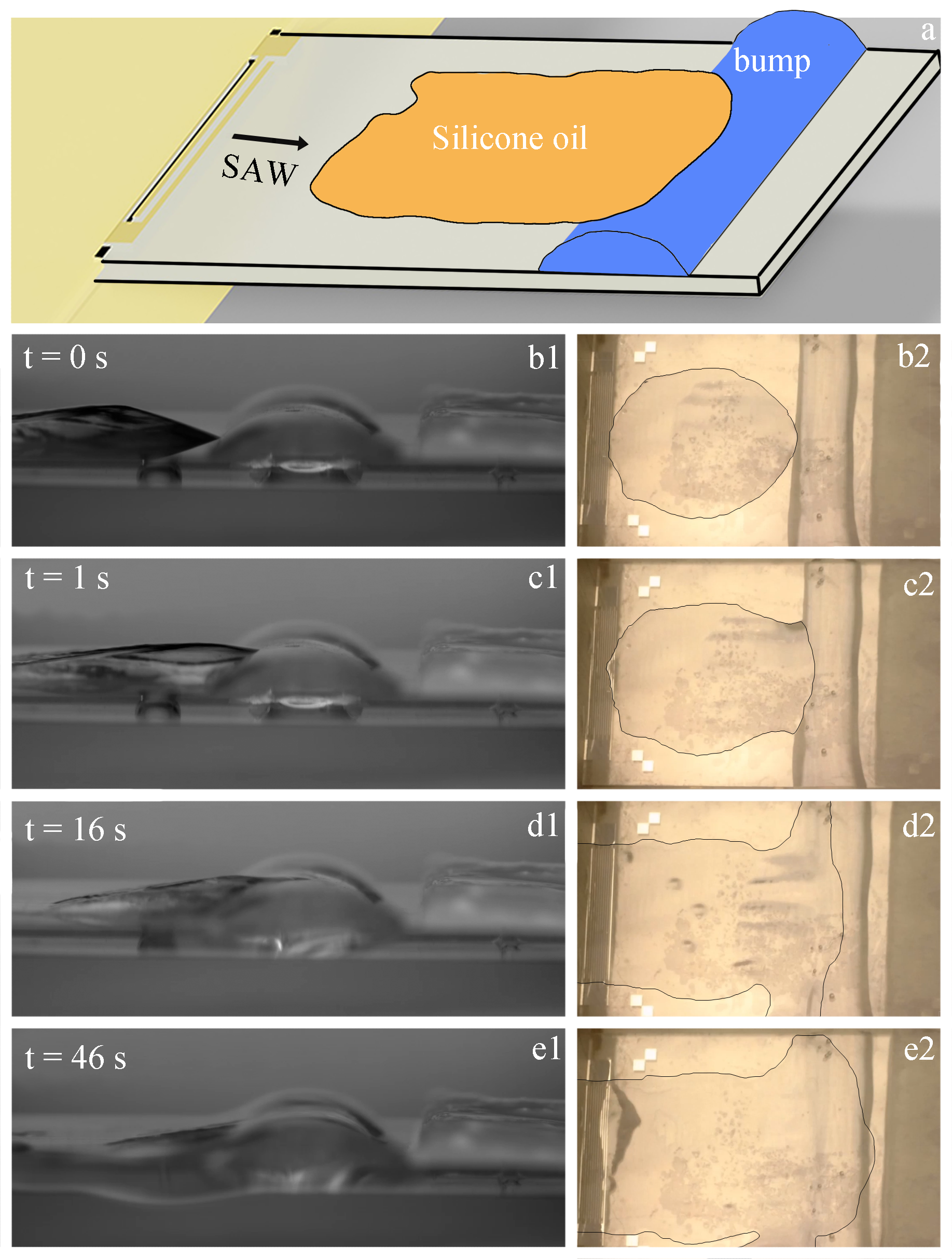}
    \caption{%{\color{red} ** Need to describe (a) too **}
    (a) A sketch of the oil covering bump obstacle and the (b1–e1) side view and (b2–e2) corresponding top view snapshots of a SAW driven silicone oil film climbing over and coating a bump; the oil film outline is marked by a dark, thin line. The snapshot sequence commences at $t=0$ (b1, b2) when the oil front reaches the rear of the bump. The oil then climbs over it under the influence of SAW (c1--d1; c2--d2), and ultimately reaches the bump front after 46 seconds (e1, e2). (Full experimental videos available upon request.)} 
    \label{fig:exp_bump}
\end{figure}

\begin{figure}[htp]
    \centering
    \subfigure[]{
\includegraphics[width=0.46\linewidth]{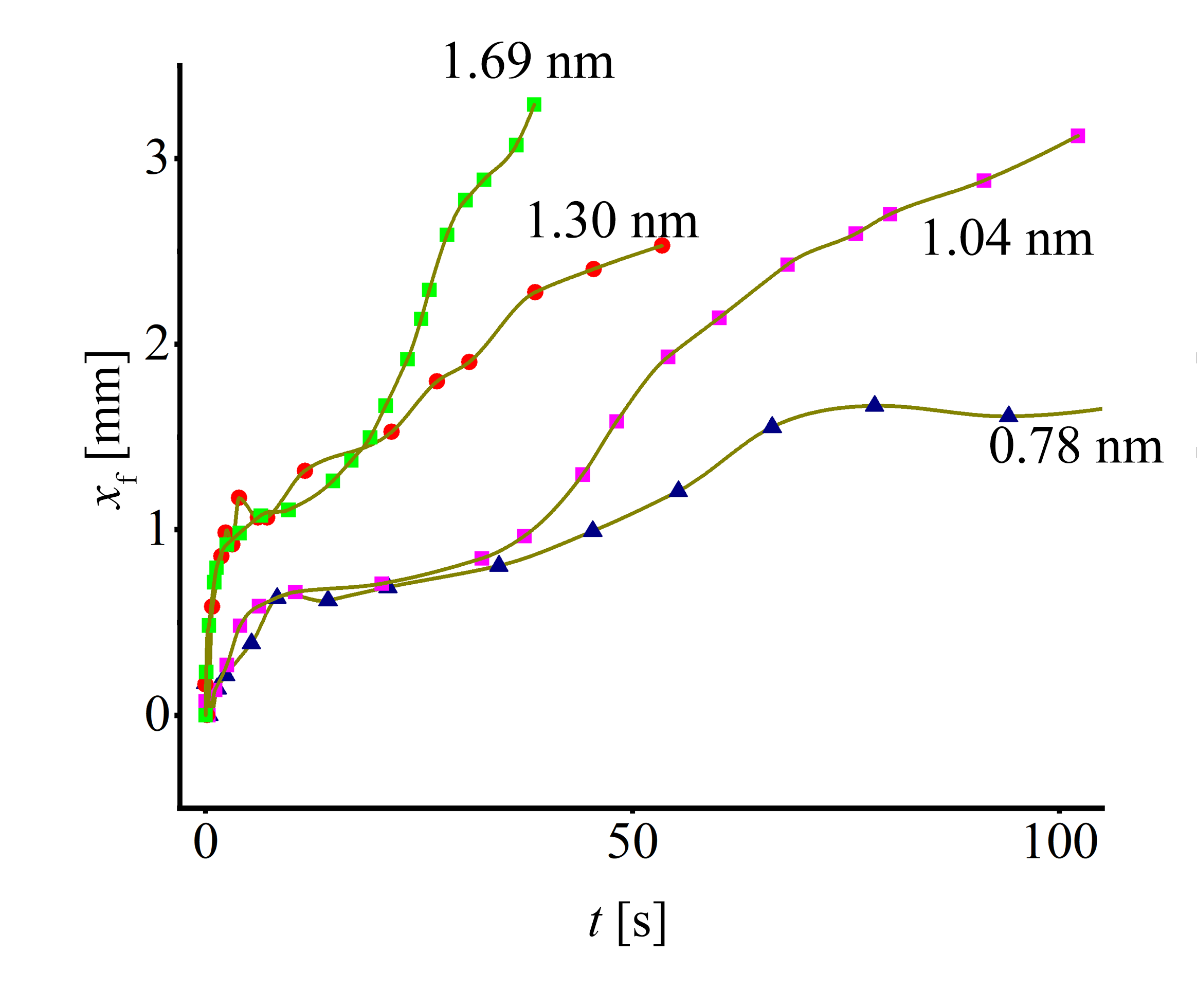}}
 \subfigure[]{
\includegraphics[width=0.46\linewidth]{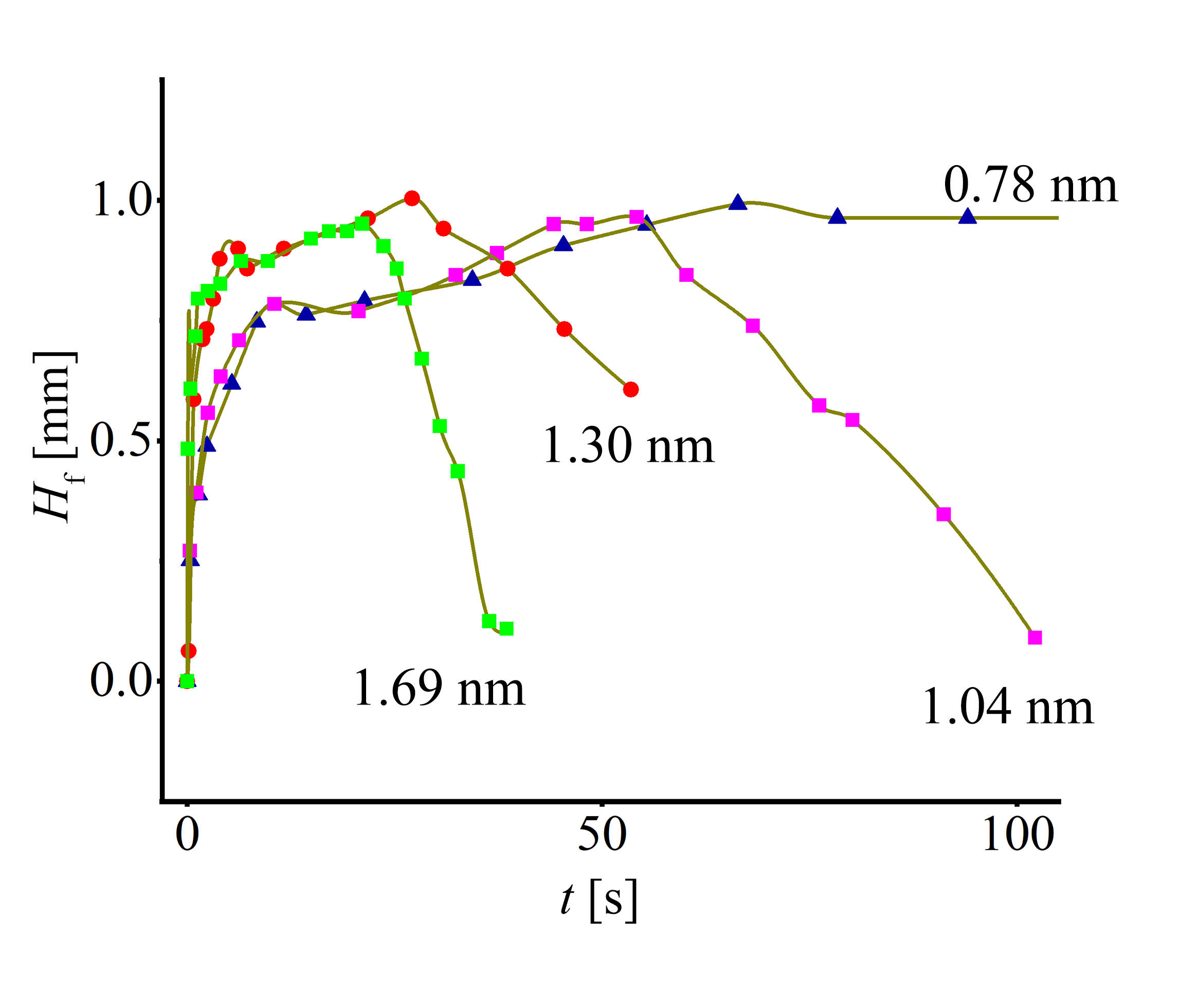}}
    \caption{(a) Front position, $X_{\rm f}$, measured relative to the rear (left edge) of the bump, and (b) front height, $H_{\rm f}$, as functions of time, for a bump of height $h_{\rm o}=0.92$~mm. Time is measured from the moment the oil reaches the rear of the bump. The SAW amplitude, $A_n$, is varied from $0.78$~nm to $1.69$~nm. 
    %{\color{red} ** Before you said due to spatial resolution -- which is it?? **}. 
    Symbols are experimental data and the connecting curves are guides for the eye. These measurements have a $5\%$ error due to the limited angle and resolution of the side view camera.
     %   \note[LK]{the label 200 is partially cut off.}\note[YF]{Fixed. limited angle and resolution both contribute to the error.}
    }
    \label{fig:exp_front_P4}\end{figure}

% \begin{figure}[htp]
%     \centering
%      \subfigure[]{
% \includegraphics[width=0.46\linewidth]{SimplifiedModel/figs_exp/Oil_obstacle_P7_a.png}}
%  \subfigure[]{
% \includegraphics[width=0.46\linewidth]{SimplifiedModel/figs_exp/Oil_obstacle_P7_b.png}}
% \caption{Same as Fig.~\ref{fig:exp_front_P4} for $h_{\rm o}=1.1$~mm; here $A_n$ is varied from 0.52 nm to 1.69 nm.
% %\note[LK]{The time range in parts (a) and (b) should be the same.  }\note[YF]{Fixed.}
% }
%     \label{fig:exp_front_P7}
% \end{figure}

In a typical experiment, the traveling oil front first reaches the rear of the bump and coats it under SAW power. We use the side camera video images to extract the contact point position over time and obtain results such as those in Fig. \ref{fig:exp_front_P4} (results shown for a bump of height $h_{\rm o}=0.92$~mm and for several different SAW amplitudes). Here, we capture both the horizontal position and the height of the front: part (a) displays the motion in the $x$--direction (the direction of SAW propagation), while part (b) illustrates motion in the out-of-the-plane $z$--direction. Generally, we observe in Fig.~\ref{fig:exp_front_P4} that as the applied SAW amplitude increases, so does the coating velocity (given by the slope of the curves in part (a)). Together, parts (a) and (b) show that initially the front of the coating film rises quickly up the bump under the influence of capillarity (for approximately 5 seconds, taking the $A_n=1.04$ nm case as an example). Then the velocity of the coating front decreases; see the period $5~-~45$ seconds for the $A_n=1.04$~nm case in Fig.~\ref{fig:exp_front_P4}(a).  Following this decrease, the thickness of the oil film builds up; as this happens and the oil passes the peak of the bump obstacle (evident from the turning point of the corresponding curve in part (b)), the coating velocity increases again: see the period from 45 seconds to the end of the experiment for the $A_n=1.04$ nm case in Fig. \ref{fig:exp_front_P4}. (Full experimental videos available upon request.)  The trends across various SAW forcing amplitudes and bump geometries are also considered via numerical simulations, discussed later in Sec.~\ref{sec:numerical_results}.

\begin{figure}[htp]
    \centering
\subfigure[$h_{\rm o}=0.26$ mm, $w_{\rm o}=1.8$ mm ]{
\includegraphics[width=0.46\linewidth]{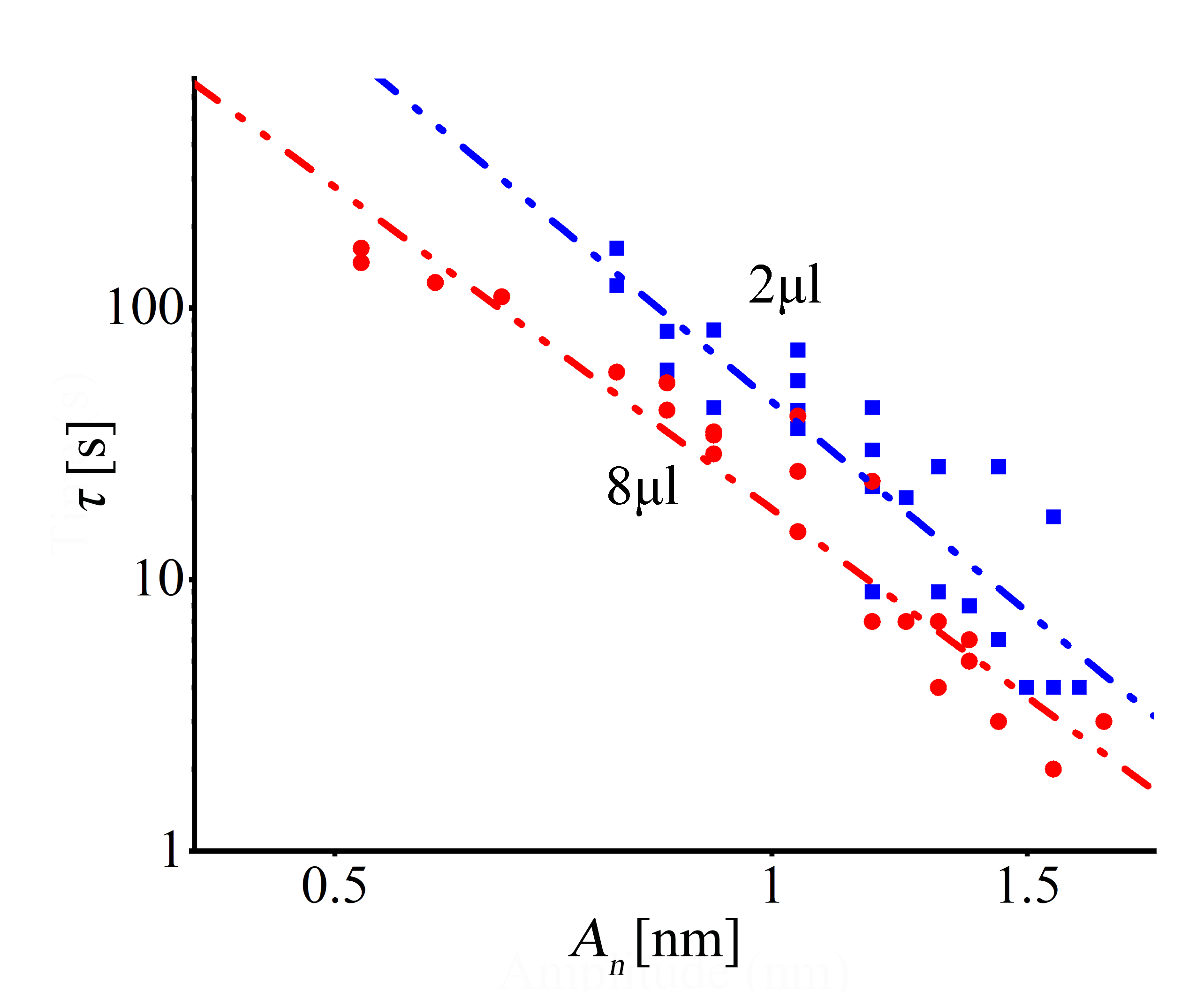}}
%  \subfigure[$h_{\rm o}=0.33$ mm, $w_{\rm o}=2.8$ mm ]{
% \includegraphics[width=0.46\linewidth]{SimplifiedModel/figs_exp/TIME_BUMP_b.png}}
\subfigure[$h_{\rm o}=0.65$ mm, $w_{\rm o}=2.5$ mm ]{
\includegraphics[width=0.46\linewidth]{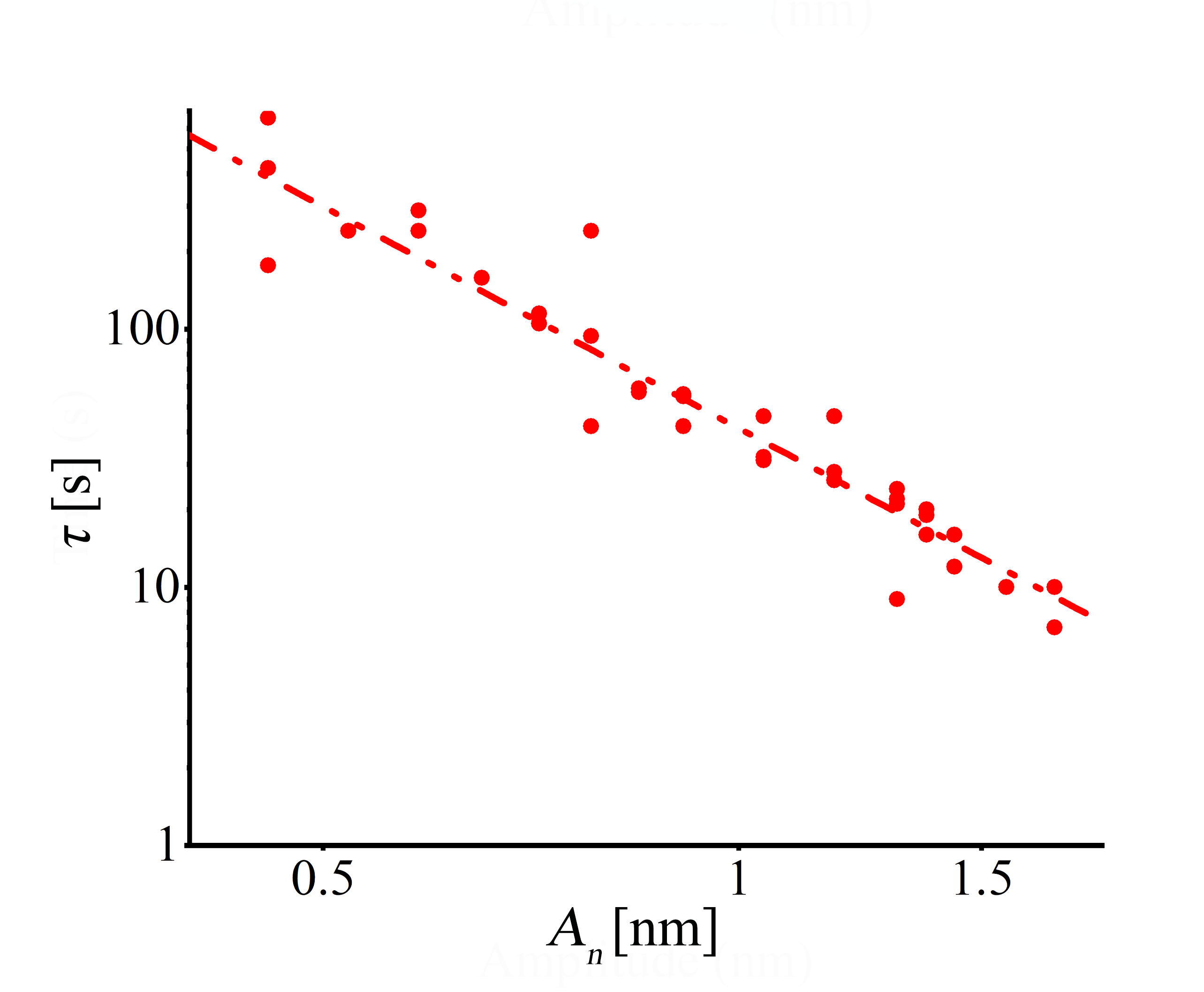}}
 \subfigure[$h_{\rm o}=0.92$ mm, $w_{\rm o}=1.8$ mm ]{
\includegraphics[width=0.46\linewidth]{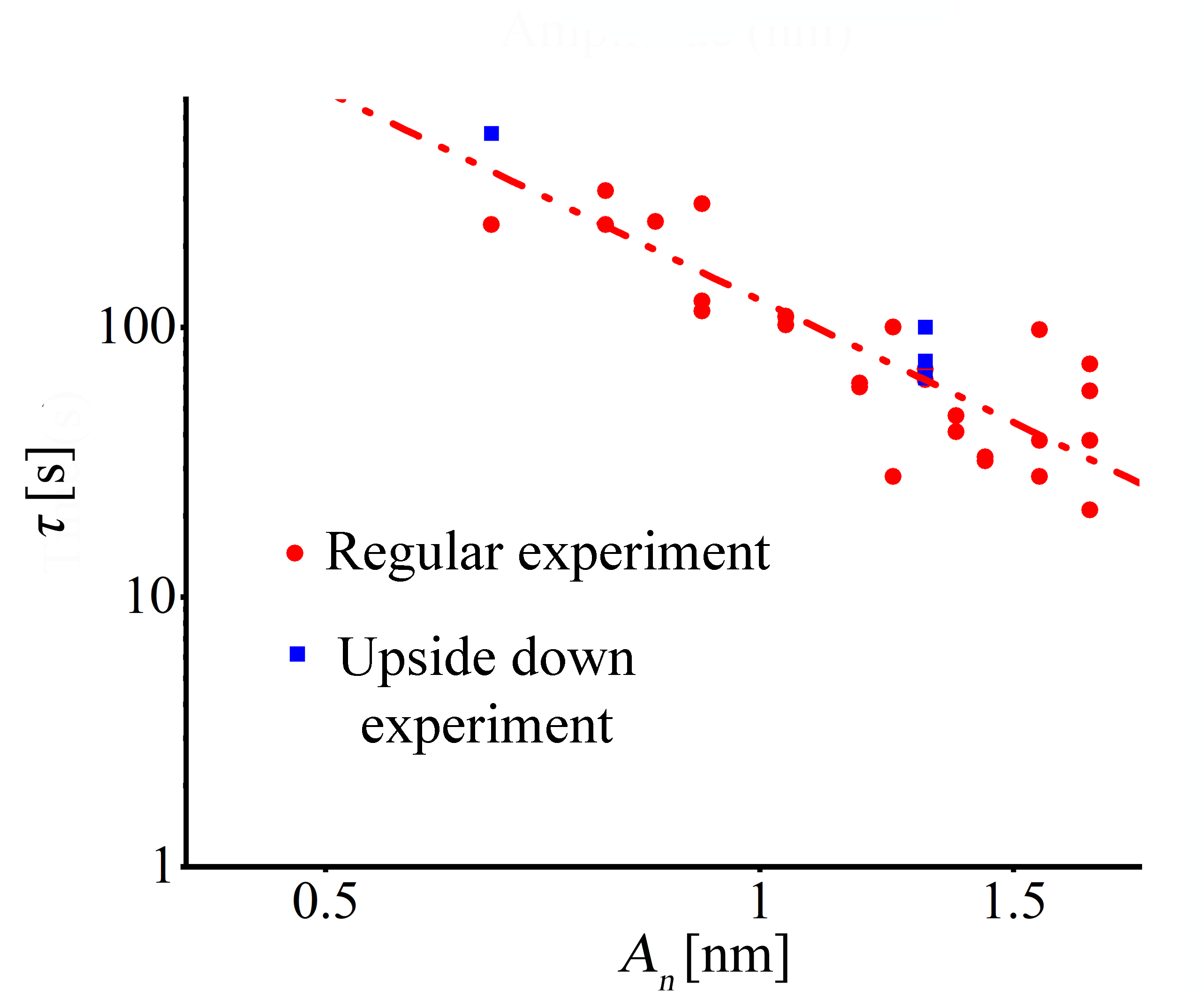}}
    \caption{Time $\tau$ to traverse the bump obstacle as a function of $A_n$ for different bump heights: (a) $h_{\rm o}=0.26$~mm, (b) $h_{\rm o}=0.65$~mm, and (c) $h_{\rm o}=0.92$~mm. 
        The red dots correspond to a drop of volume $V_d=8~\mu$l. The blue dots in (a) are for a drop of volume $V_d=2~\mu$l, while in (c) they stand for the upside down experiment. Climbing time is the time elapsed from when the drop front touches the rear of the bump to when it reaches the front.    }
    \label{fig:exp_time_bump}
\end{figure}

We also measure the time $\tau$ taken for silicone oil to fully traverse bumps of different heights as a function of SAW amplitude $A_n$ under a range of conditions; see the log-log plots in Fig.~\ref{fig:exp_time_bump}. As expected, we find that larger $A_n$ results in shorter coating times: in fact, the logarithm of the coating times varies linearly with the logarithm of $A_n$, suggesting that the two quantities are related by a power law. Regarding the influence of the bump height on the results, we observe in Fig.~\ref{fig:exp_time_bump} a general trend that larger heights lead to longer times, although there is a significant spread of the resulting times; once again, a better understanding of the trends of the results can be achieved by numerical simulations presented later in the text.

To further enhance our understanding of the coating process, we also show in Fig. \ref{fig:exp_time_bump}(a) the results of experiments carried out using a smaller volume, 2 $\mu$l, of silicone oil (aside from the usual 8 $\mu $l). The results show that, in general, it takes longer for this smaller volume of silicone oil to coat the bump. Although the slopes of the best-fit lines on the log-log plots for the two sets of experiments are different, as $A_n$ increases (approaching 1.5 nm), the time required for the 2 $\mu$l  oil to climb over the bump approaches that of the 8 $\mu$l  case. Therefore, when the SAW amplitude is sufficiently high, the coating effect is less sensitive to the initial volume of the oil film. The observed power-law relation between $A_n$ and coating time persists over a range of bump heights; see Figs.~\ref{fig:exp_time_bump}(a-c). 

Finally, in addition to the experiments reported so far, we also conducted the experiments in an "upside down" configuration.  The results of such experiments are shown in Fig.~\ref{fig:exp_time_bump}(c) alongside the results for the regular configuration: the results confirm that gravity plays only a negligible role.
%: the dominant mechanisms that govern the coating in these experiments are capillary and SAW-induced stresses. 

\section{Theory} \label{sec:model}
We base our model of the experimental system described in Sec.~\ref{sec:experiment} on the two-dimensional (2D) framework developed by Fasano et al.\cite{Li2024}, who described the spreading of silicone oil films driven by surface acoustic waves (SAWs) on a flat substrate. In that work asymptotic analysis, based on assumptions that the oil film free surface has small slope (the long-wave or lubrication approximation) and that the acoustic Mach number is small, was used to derive a fourth order nonlinear partial differential equation governing the film height, with a SAW-induced stress term driving flow. Here, we build on that model to account for the presence of topographical obstacles made of PDMS, modifying all forcing terms in the resulting ``thin-film'' governing equation, consistent with existing literature where available. Again, the modeling is simplified by considering a 2D viscous oil film driven by a SAW propagating along the substrate and under the obstacle (also 2D), in the positive $x$-direction. In what follows, we use the similar acoustic properties of the cured PDMS and liquid silicone oil to assume that the SAW attenuates in the same way under the PDMS obstacle and under the silicone oil. We further note that the extension of the model to include a topographical obstacle does not modify the leading-order quiescent or first-order oscillatory flows in the asymptotic expansion of the oil film from those derived earlier~\cite{Li2024}, but it does modify the second-order streaming flow that drives the oil film dynamics.

\subsection{Nondimensionalization and scaling}
\label{sec:nd}
We assume that the viscous silicone oil film has a density $\rho$ and dynamic viscosity $\mu$. The surface tension of the oil-air interface is denoted by $\gamma$. The acoustic field is characterized by the maximum vertical displacement amplitude of the solid substrate due to the SAW, $A$, and the angular frequency of the SAW, $\omega=2\pi f$. 
We carry out a standard nondimensionalization to highlight the balances between the considered forces in the model. Let $L=1$~mm be a characteristic length scale based on the acoustic attenuation length under the oil film~\cite{Li2024}; all lengths, including the film height, $h(x,t)$, are scaled with $L$. Time is scaled with $T=3\mu L/\gamma$ to balance viscous and capillary forces in the dimensionless governing equations. Pressure is scaled with $P=\gamma/L$ so that the curvature term enters the governing equations with unit coefficient. With gravitational acceleration $g$ due to gravity acting in the $z$ direction, the nondimensional groups used in the governing equations are defined as
\begin{equation}
    \text{Bo} = \frac{\rho_0gL^2}{\gamma}, \qquad \mathcal{S}=\frac{P_0L}{\gamma}=\frac{\rho_0A^2\omega^2L}{\gamma},
\end{equation}
where $\text{Bo}$ is the Bond number, which measures the relative strength of gravity and surface tension and $\mathcal{S}$ measures the relative strength of the SAW-induced stress and surface tension. Here, $\rho_0$ is the constant, leading-order term in the asymptotic expansion of the oil film density in the small acoustic Mach number (see Appendix \ref{app:SpreadingModel} for further details). Additionally, $P_0$ is a SAW-induced stress scale, with units of pressure, that gives the characteristic magnitude of the time-averaged Reynolds stresses produced by the first-order ultrasonic field that diffract (leak) off the SAW (see Appendix \ref{app:SpreadingModel} for further details, specifically the force components in Eqs.~\eqref{eq:Fsx}-\eqref{eq:Fsz}). We note that in what follows we will use the non-dimensional formulation to present the governing equations, but will show results in terms of physical variables to facilitate comparison with experiment. 

We comment here also on the relation between the experimentally measured SAW amplitude, $A_n$, and the theoretical one, $A$, used in Appendix \ref{app:SpreadingModel} to derive the model, which we use in accordance with \cite{Li2024}. Although we expect $A\propto A_n$, it is not clear that these amplitudes are identical, as discussed in some detail by Royer \& Dieulesaint\cite{Royer1996}. A detailed analysis to determine the exact relation between $A$ and $A_n$ is beyond the scope of the present work; we will comment further regarding this relation  when discussing the comparison between experimental and theoretical results in Section \ref{sec:discussion}.

\subsection{Governing Equations}
On a flat substrate without topographical obstacles, the nondimensional evolution equation derived in \cite{Li2024} can be written as 
\begin{equation}
    \frac{\partial h}{\partial t}+\frac{\partial}{\partial x}\bigg[-h^3 \frac{\partial \mathcal{P}}{\partial x} - \mathcal{C} \frac{3\mathcal{S}}{8 K_z^4}\psi(x,h)\Big(2 K_z^2 h^2-1+e^{2 K_z h}(1-2 K_z h) \Big) \bigg]=0, \label{eq:dhdtadim}
\end{equation}
where the nondimensional effective pressure is (with $'$ representing $\partial/\partial x$)
\begin{equation}
    \mathcal P = -h^{\prime \prime} + {\rm Bo}\, h  + \frac{\mathcal{S} C_z}{2 K_z} \psi( x,h).
    \label{eq:Pad}
\end{equation}
The first two terms in $\mathcal{P}$ represent the effects of capillarity and gravity, respectively, while the final term describes contributions arising from the acoustic field in the oil, with $\psi(x,h)$ given by 
\begin{equation}
    \psi(x,h) = \left\{ 
    \begin{array}{ll}
    1, & x< x_{\rm r}^\ast(t)\, , \\
    e^{-2 \left[ k_{\rm s,\rm i} (x - x_{\rm r}^\ast(t)) + K_z  \left( h - h^\ast \right) \right]}, & x_{\rm r}^\ast(t) \leq x \leq x_{\rm f}^\ast(t)\, , \\
    e^{-2 k_{\rm s,\rm i} \left[ x_{\rm f}^\ast(t) - x_{\rm r}^\ast(t) \right]}, & x > x_{\rm f}^\ast(t) \, .
    \end{array}
    \right.
    \label{eq:phis_hcut}
\end{equation}
Further acoustic parameters that appear in the above model include: a measure of the deviation of the SAW-induced stress from conservative form, $\mathcal{C}$; the attenuation coefficient of the leaky SAW in the liquid including viscous effects in the $z$-direction, $K_z$; the attenuation coefficient of the leaky SAW in the $x$-direction, $k_{\rm s,i}$; and the coefficients of the acoustic body force in the $x$ and $z$ directions, $C_x$ and $C_z$, respectively. See Appendix \ref{app:SpreadingModel} for full definitions and further details on how these parameters arise in the mathematical model. In simulations, we assume that SAW attenuation starts at the rear cutoff line location, $x_{\rm r}^\ast$, and ends at the front cutoff line location, $x_{\rm f}^\ast$, defined by the points where the film height reaches a specified cutoff thickness, $h(x_{\rm r}^\ast,t)=h(x_{\rm f}^\ast,t)=h^\ast$ (in practice, each of these points is uniquely defined). The value of the critical cutoff thickness at which the SAW begins attenuating, $h^\ast$, is chosen based on the thickness below which mass transport is dominated by other mechanisms different from that of Eckhart streaming, such as Rayleigh streaming and acoustic radiation pressure~\cite{Rezk12,Rezk14}. This thickness turns out to be  $h^\ast\approx(\lambda_{\text{oil}}/4)/L=17~\mu$m$/L$.  We note that the SAW contribution to the governing equation (\ref{eq:dhdtadim}) has both conservative and non--conservative parts: the former corresponds to $\mathcal{P}$ in the last term on the right hand side of Eq.~(\ref{eq:Pad}), and the latter to the final term (the $\mathcal{C}$-contribution) in Eq.~(\ref{eq:dhdtadim}).

When a topographical PDMS obstacle of height profile $s(x)$ is introduced, the oil film thickness $h (x,t)$ is measured relative to the obstacle, so that the total height of the free surface above the substrate is $H=h+s$. The governing equation is then modified to
\begin{eqnarray}
    \pder[h]{t} + \pder{x}\bigg[-h^3 \pder[\mathcal{P}]{x}
    -\mathcal{C}\frac{3\mathcal{S}}{8K_z^4}\psi(x,h+s)\Big(2K_z^2h^2-1+e^{2K_zh}(1-2K_zh)\Big)\bigg] = 0 , \label{eq:ndGovEqMain}
\end{eqnarray}
with effective pressure
\begin{equation}
    \mathcal{P} = -\Big(h''+s''\Big)+ \text{Bo}\, (h+s)+\frac{\mathcal{S}C_z}{2K_z}\psi(x,h+s). \label{eq:ndEffPMain}
\end{equation}
Here, the capillary contribution to $\mathcal{P}$ in Eq.~\eqref{eq:ndEffPMain} is modified through the obstacle's contribution to the curvature of the oil film, consistent with the work of Park and Kumar~\cite{park2017droplet}. Similarly, the gravitational term accounts for the increased height of the film due to the obstacle (see also Stillwagon and Larson~\cite{stillwagon1988fundamentals}). The SAW-induced stress is affected by the obstacle only via the vertical attenuation, through its inclusion in the second argument of $\psi$. This is a result of our assumption that the solid PDMS obstacle has the same attenuation coefficients as the silicone oil; for other fluids (or obstacles), this assumption may not hold. For the same reason, we modify the definition of the rear and front cutoff line locations to account for the additional height of the solid obstacle beneath, defining $x_{\rm r}^\ast,x_{\rm f}^\ast$ as the points such that $h(x_{\rm r}^*,t)+s(x_{\rm r}^*)=h(x_{\rm f}^*,t)+s(x_{\rm f}^*)=h^\ast$. 

In the absence of an obstacle ($s=0$), Eqs.~\eqref{eq:ndGovEqMain} and~\eqref{eq:ndEffPMain} reduce to the flat-substrate model. Additionally, without acoustics ($\mathcal{S}=0$), our model is in agreement with that of Park and Kumar\cite{park2017droplet}. Before discussing numerical results, we point out that Eq.~\eqref{eq:ndGovEqMain} is of the form $h_t+(vh)_x=0$ where $v$ is the thickness-averaged horizontal velocity component of the silicone oil, given by
\begin{eqnarray}
    v = -h^2\frac{\partial\mathcal{P}}{\partial x} -\mathcal{C} \frac{3\mathcal{S}}{8hK_z^4}\psi(x,h+s)\bigg(2K_z^2h^2-1+e^{2K_zh}(1-2K_zh)\bigg).  
    \label{eq:vMain}
\end{eqnarray}
See Appendix \ref{app:lub} for the detailed long-wave reduction leading to Eq. \eqref{eq:ndGovEqMain}. 

\section{Results} \label{sec:numerical_results}
 
This section is structured as follows: Sec. \ref{sec:numEqs} introduces the initial and boundary conditions relevant to the experiment needed to close the model; then Secs. \ref{sec:ramp} and \ref{sec:bump} show simulations conducted with a ramp obstacle and a bump obstacle, respectively. In each of Secs. \ref{sec:ramp} and \ref{sec:bump}, we first present a representative simulation, then we show how the dynamics changes as we vary the SAW amplitude, obstacle dimensions, or oil drop volume. Simulations are compared with experiments in selected figures; to facilitate this, all figures are plotted in dimensional units. The SAW frequency is always fixed at $20$ MHz; only the maximum vertical displacement amplitude of the solid substrate due to the SAW, $A$ (measured in nanometers) is varied. The dimensionless volume is written as $\mathcal{V}_{\rm d} = V_d/L^3$ and we consider that the standard drop volume is  $V_d= 8 \,\mu l$, except where otherwise stated.

\subsection{Initial and Boundary Conditions} \label{sec:numEqs}
In our simulations, we solve Eq.~\eqref{eq:ndGovEqMain} with effective pressure given by Eq.~\eqref{eq:ndEffPMain}; the main ideas of the implementation in COMSOL\texttrademark~are similar to those described in Appendix C of \cite{Li2024} for the dynamics on a flat substrate. The film is initialized with a two-dimensional parabolic profile of constant dimensionless area $\mathcal{A}_{\rm d}$, corresponding to a cross-sectional slice of a three-dimensional parabolic cap, given by 
\begin{equation}
\label{eq:Ad}
{\cal A}_{\rm d} = 2\int_0^{r_{\rm d}} h_{\rm d}  \left[ 1- \left(\frac{x}{r_{\rm d}}\right)^2 \right] \,dx =
\frac{4}{3} h_{\rm d}  r_{\rm d}, 
\end{equation}
where $r_{\rm d}$ is the radius of the oil drop and $h_{\rm d}$ its maximum height (both dimensionless). This height $h_{\rm d}$ is obtained by using $h_{\rm d}=2\mathcal{V}_{\rm d}/(\pi r_{\rm d}^2)$, which follows from the formula for the volume $\mathcal{V}_{\rm d}$ of a $3$D parabolic cap of known base radius, $r_{\rm d}$. Here, we consider $r_{\rm d}=3.2$ as a good approximation of the initial drop radius, since it is placed with a pipette in the experiments, and determine its height $h_{\rm d}$ for given drop volume. Thus,  all parameters are chosen to match the dimensional drop volumes used in experiments, but are here expressed in dimensionless form. The initial oil drop is centered at $x=x_{\rm d}$ with rear and and front contact line positions at $x_{\rm r}(0)=x_{\rm d}-r_{\rm d}$ and $x_{\rm f}(0)=x_{\rm d}+r_{\rm d}$. To regularize the moving contact line, we impose a thin precursor film of thickness $h_{\rm p}\ll h_{\rm d}$, and we discretize the domain with grid spacing $\Delta x=h_{\rm p}$ to ensure numerical convergence \cite{DK_jcp02}. In all results shown, we use $h_{\rm p}=0.01$. The initial condition for the drop profile is then
\begin{equation}
        \label{eq:hinit2D}
   h(x,0)= \left\{ 
        \begin{array}{ll}
             h_{\rm p} ,&  x <  x_{\rm r}(0) ,\\
            \left( h_{\rm d} -  h_{\rm p} \right) \left[ 1- \left(\frac{ x- x_{\rm d}}{ r_{\rm d}}\right)^2 \right] +  h_{\rm p} ,& 
            x_{\rm r}(0)\leq x \leq  x_{\rm f}(0),\\ 
             h_{\rm p} ,& x >   x_{\rm f}(0), 
        \end{array} 
        \right.
\end{equation}
ensuring that the drop thickness profile is piecewise continuous and has the prescribed area $\mathcal{A}_{\rm d}$.

At the lateral domain boundaries, we fix the height to the precursor thickness via Dirichlet boundary conditions ($h=h_{\rm p}$) and impose vanishing Neumann boundary conditions on $h$ ($h'=0$). These boundary conditions ensure that the simulation captures the dynamics of the moving film without introducing spurious mass loss or gain from either domain end. The computational domain is chosen large enough so that the film does not interact with the boundaries; if the film were to reach a domain edge, the imposed Dirichlet condition would no longer be valid.

\subsection{Spreading on a Ramp} \label{sec:ramp}
We first investigate spreading over a ramp-shaped obstacle, designed to mimic the experimental setup shown in Fig. \ref{fig:ramp}. Since the silicone oil film reaches the top of the ramp in only a small number of the experimental realizations (see Fig.~\ref{fig:exp_height}), we simplify to an infinite ramp profile in our simulations, ensuring that a finite equilibrium film climbing height on the ramp can always be found. The ramp profile is specified as a smooth linear incline of height $h_{\rm o}$ and width $w_{\rm o}$, beginning at position $x=x_{\rm o}$:
\begin{equation}
s(x)= \left\{
\begin{array}{lll}
0,\quad & x <x_{\rm o},  \\
\frac{h_{\rm o}}{w_{\rm o}}(x-x_{\rm o}),\quad & x > x_{\rm o},   \end{array}
\right. \label{eq:ramp}
\end{equation}
with a smoothing layer near $x=x_{\rm o}$ of width $0.5$ to ensure continuous first and second derivatives. Unless otherwise stated, we use $h_{\rm o}=3.2$, $w_{\rm o}=3.5$ (to match the slope of the experimental ramp), and $x_{\rm o}=20$ (so that the entire film is to the left of the ramp initially). Representative snapshots of the simulation for $A=4$~nm are shown in Fig.~\ref{fig:ramp_Snapshots}, where the evolution of the film profile and the horizontal component of the thickness-averaged velocity (given by Eq.~\eqref{eq:vMain}) are shown along with the ramp obstacle, at various times indicated in the figure caption. The height and velocity profiles are made dimensional via scaling by $L$ and $L/T$, respectively. Before reaching the obstacle, the film translates steadily due to the balance of capillarity, gravity, and SAW-induced stress, similar to the flat-substrate dynamics described by Fasano et al.~\cite{Li2024}. Once the film front reaches the ramp, its dynamics change markedly. As the film front encounters the increasing substrate height, the free surface ($h+s$) is forced to bend more sharply, increasing the local curvature and thus the capillary (surface tension) pressure gradient at the front. This capillary driving overcomes the additional gravitational resistance associated with the increased height of the ramp, leading to a transient speedup of the fluid front. Simultaneously, the rear of the film continues moving at a quasi-constant speed, unaffected by the effects at the front caused by the obstacle, leading to a temporary increase in the horizontal width of the film. This speedup effect of the front is fleeting, however, as evidenced by the sharp decrease in the velocity near the oil front in Fig.~\ref{fig:ramp_Snapshots}(b) compared to that in Fig.~\ref{fig:ramp_Snapshots}(a). Rather than accumulating fluid at the rear though, the film gradually redistributes mass along the incline, flattening with respect to the ramp obstacle and extending as it climbs [Fig.~\ref{fig:ramp_Snapshots}(b)-(c)]. During this stage, the rear contact line is still driven forward at the quasi-constant speed (with which it translates in the case of the flat substrate) by the acoustic stress. This causes the horizontal width to contract, since the rear is now moving faster than the front. This contraction leads to a second speedup of the front, due to the increased SAW-induced stress felt there as a result of the decreased horizontal attenuation of the force over the shorter drop length. Eventually, as shown in Fig.~\ref{fig:ramp_Snapshots}(d), capillary and gravitational forces balance the attenuated SAW-induced stress and the film reaches a steady-state position on the ramp. The slight backflow (negative velocity) seen in Fig.~\ref{fig:ramp_Snapshots}(b) and Fig.~\ref{fig:ramp_Snapshots}(d) behind the bulk is to satisfy conservation of mass in the precursor below the cutoff thickness at which attenuation turns on (i.e. in areas where $H=h+s<h^\ast$). This is consistent with the dynamics seen on a flat substrate \cite{Li2024}. Full animations are available upon request for various values of $A$. 

\begin{figure}
\centering
%\subfigure[]{
%\includegraphics[width=0.48\linewidth]{newFigs/Fig6a_1_A130_t8.png}}
%\subfigure[]{
%\includegraphics[width=0.48\linewidth]{newFigs/Fig6b_A130_t3.png}}
\subfigure[~$t=1$ s]{
\includegraphics[width=0.48\linewidth]{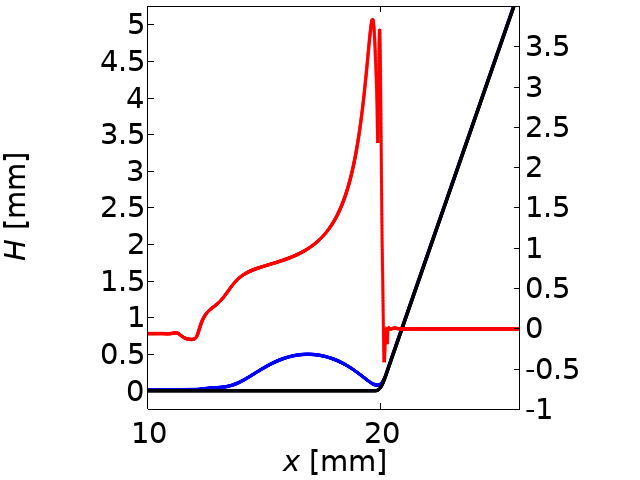}}
\subfigure[~$t=2.5$ s]{
\includegraphics[width=0.48\linewidth]{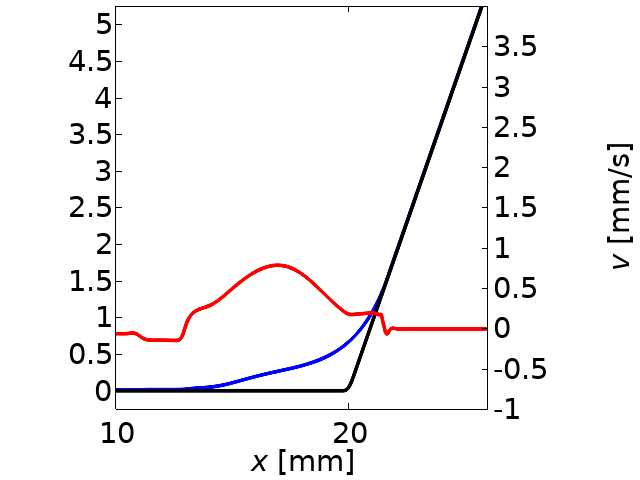}}
\subfigure[~$t=12.5$ s]{
\includegraphics[width=0.48\linewidth]{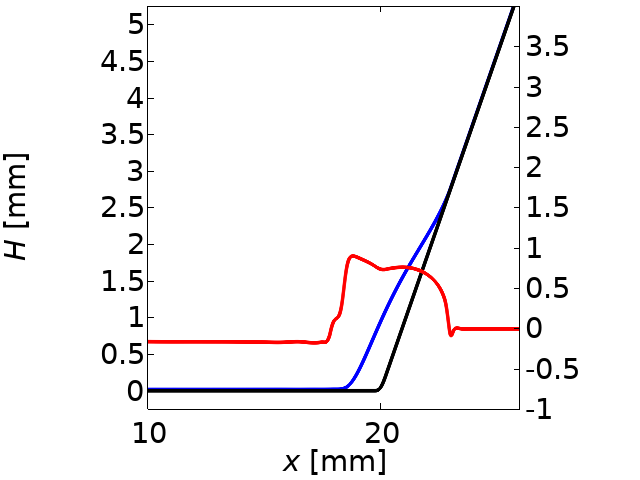}}
\subfigure[~$t=50$ s]{
\includegraphics[width=0.48\linewidth]{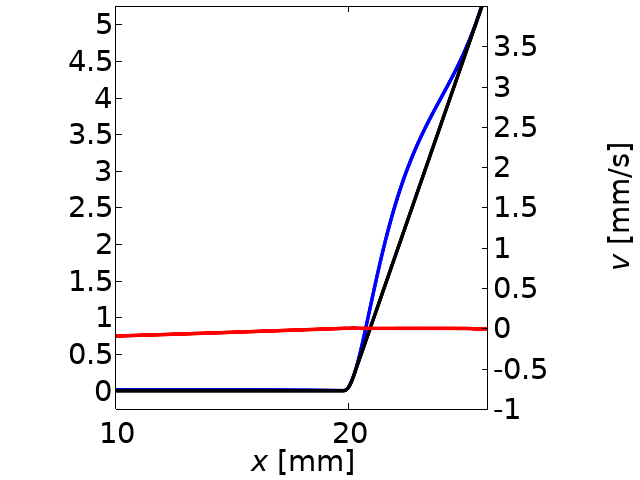}}
\caption{Evolution of the thickness profile as the film climbs the ramp, for $A = 4$ nm at different times. The black solid line denotes the underlying ramp obstacle. The {blue} solid line represents the height profile of the liquid film at the current time. The {red} line indicates the thickness-averaged horizontal velocity component, $v$, given by Eq.~\eqref{eq:vMain}, dimensionalized by the scale $L/T$. Full simulation videos available upon request for a range of $A$ values.
}
\label{fig:ramp_Snapshots}
\end{figure}

The dependence of climbing behavior on forcing amplitude is summarized in Fig.~\ref{fig:climbingHeights_FinalTimes}(a), which shows the front climbing height, $H_{\rm f}=s(x_{\rm f}(t))$ where the front contact line is determined as the point $x_{\rm f}$ where $h(x_{\rm f})=h^\ast$, as a function of time. For weak SAW-induced stress ($A\lesssim1$~nm), the film front only climbs a short distance up the ramp because the rear contact line spreads backward due to capillary forces dominating the SAW-induced stress in this region of the film. This is a significant effect because it causes the attenuation of the acoustic force at the fluid front to increase in time as the rear drifts further backward. Although $x_{\rm r}^\ast$ continues to drift left, the attenuation factor $\psi$ depends on both horizontal path length and local thickness. As the rear moves backward, the film simultaneously thins and levels on the ramp, reducing the vertical attenuation and thereby compensating in part for the increase in horizontal attenuation. Consequently, the location of the $h=h^\ast$ contour used to define the front can remain nearly fixed, yielding a stationary climbing height even while the rear cutoff continues to evolve in the opposite direction as the propagating wave. 
For intermediate SAW amplitude ($1~{\rm nm}\lesssim A\lesssim2~\rm{nm}$), the film exhibits mixed behavior between the weak- and strong-forcing regimes. In this range, the rear cutoff location $x_{\rm r}^\ast$ drifts backward due to capillarity as in the weak-forcing case, but this motion slows and may temporarily arrest as the SAW-induced stress becomes comparable to capillary forces. As a result, the front climbing height increases slowly in time before reaching a quasi-steady value that lies between the weak- and strong-forcing limits. This intermediate regime corresponds to a partial balance between horizontal attenuation caused by the retreating film rear and reduced vertical attenuation as the film thins and redistributes along the ramp.
For sufficiently strong SAW amplitudes ($A>2$ nm), the rear contact line advances in the direction of SAW propagation, resulting in a net forward translation of the entire film. This translation increases the climbing height by maintaining stronger acoustic forcing at the fluid front, effectively \textit{pushing} the film upward along the obstacle. The climbing height of the film increases monotonically with $A$ as expected due to this effect. Figure~\ref{fig:climbingHeights_FinalTimes}(b) shows the long-time (steady-state) height profiles for various values of the SAW amplitude $A$, with all profiles made dimensional by scaling with $L$.

\begin{figure}
    \centering
    \subfigure[]{\includegraphics[width=0.48\linewidth]{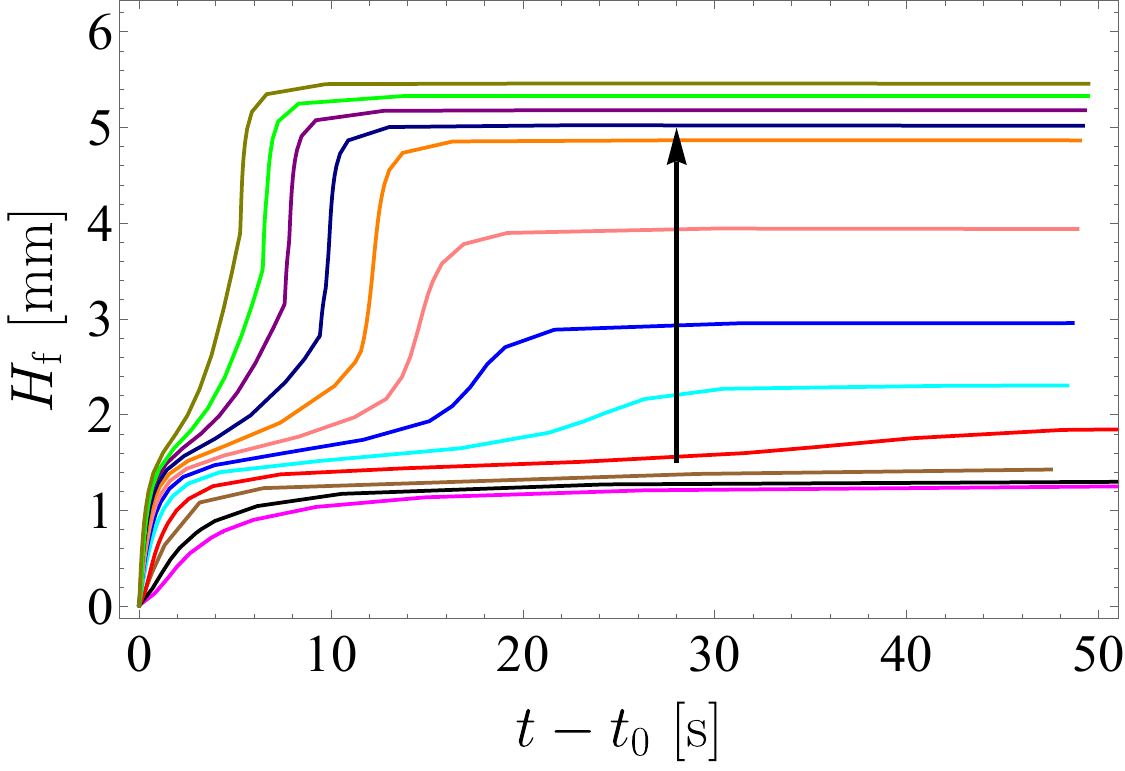}}
    \subfigure[]{\includegraphics[width=0.48\linewidth]{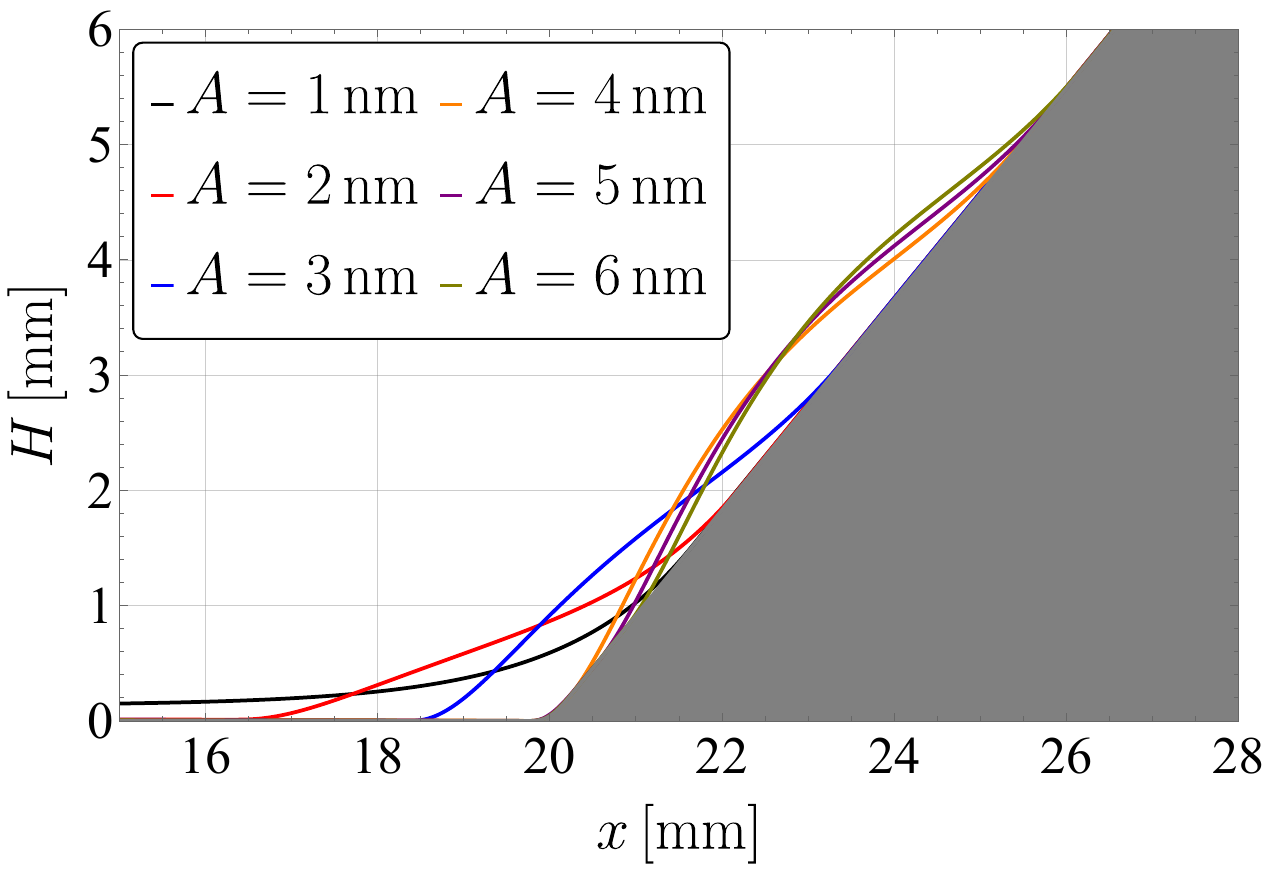}}
    \caption{(a) Climbing height of the film on the ramp, $H_{\rm f}$, as a function of time from simulations for $A=0.5$~nm (magenta) to $A=6$~nm ({olive color}) with $\Delta A=0.5$~nm increasing according to the black arrow. (b) Steady-state (at large times) height profile of the film on a ramp for different values of the SAW amplitude, $A$.
    }
    \label{fig:climbingHeights_FinalTimes}
\end{figure}

We also examine the influence of ramp geometry on results, as shown in Fig.~\ref{fig:rampSlope}. We carry out simulations for fixed SAW amplitude, $A$, varying the ramp parameters ($h_{\rm o}/w_{\rm o}$) to obtain three slopes: $31.4^\circ, 42.4^\circ$, and $53.9^\circ$. In Fig.~\ref{fig:rampSlope}(a), we plot the climbing height as a function of time (shifted so that $t=0$ corresponds to the moment the film front $x_{\rm f}$ reaches the ramp), and in Fig.~\ref{fig:rampSlope}(b), we plot the steady-state profiles reached at long times. Increasing the ramp slope influences two competing effects. On the one hand, steeper ramps produce smaller horizontal displacements of the front, reflecting stronger gravitational force along the incline and enhanced vertical acoustic attenuation. On the other hand, and somewhat counterintuitively, the maximum vertical climbing height increases with ramp slope. This behavior arises because steeper ramps reduce the horizontal extent over which the SAW attenuates before reaching a given vertical height, thereby decreasing horizontal attenuation of the acoustic forcing. In the parametric regime considered here, this reduction in horizontal attenuation more than compensates for both the increased gravitational resistance and the enhanced vertical attenuation due to larger film and obstacle thickness, resulting in higher steady-state climbing heights on steeper ramps. 

We next examine the influence of film volume on the climbing dynamics, shown in Fig. \ref{fig:rampVol}.  Simulations are performed at a fixed SAW amplitude $A=4$ nm for three film volumes, 3~$\mu$l, 5~$\mu$l and 8~$\mu$l
($\mathcal{V}_{\rm d}=3,5,8$). Figure \ref{fig:rampVol}(a) shows the climbing height as a function of time (again shifted so that $t=0$ corresponds to the instant at which the film first reaches the ramp), while Fig. \ref{fig:rampVol}(b) shows the corresponding steady-state profiles at long times. The transient climbing dynamics is nearly identical for all three volumes, and the resulting steady-state climbing heights differ only modestly. This weak dependence on volume arises from the manner in which acoustic attenuation is implemented in the model. Attenuation of the SAW-induced stress begins once the total fluid thickness reaches a fixed cutoff value $H=h+s=h^\ast$, independent of the total film volume. Consequently, for all SAW amplitudes that are able to force the film to climb fully onto the ramp, the onset of attenuation occurs at approximately the same substrate elevation, corresponding to the location $x_{\rm r}^\ast$ where $s(x_{\rm r}^\ast)\approx h^\ast-h_{\rm p}$. As a result, the spatial distribution of the SAW-induced stress beneath the film is nearly identical across the volumes considered. The modest difference in the final climbing height, therefore, primarily reflects geometric effects rather than changes in the driving force. Larger-volume films have a larger footprint and extend farther forward once the rear contact line has settled into a nearly fixed position ($x_{\rm r}^\ast\gtrsim20$ mm), leading to slightly higher measured climbing heights. The apparent increase in steady-state climbing height with volume thus does not indicate stronger acoustic driving, but instead reflects the increased spatial extent of the fluid.

\begin{figure}
    \centering
    \subfigure[]{
    \includegraphics[width=0.48\linewidth]{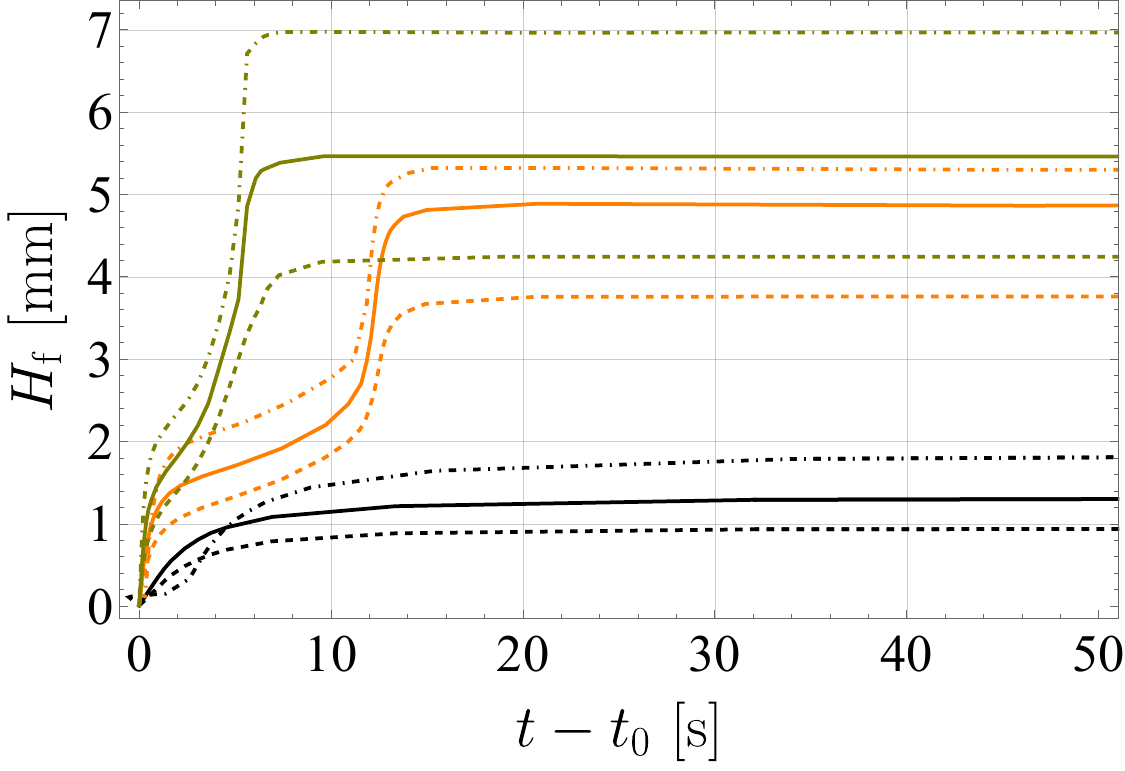}}
    \subfigure[]{
    \includegraphics[width=0.48\linewidth]{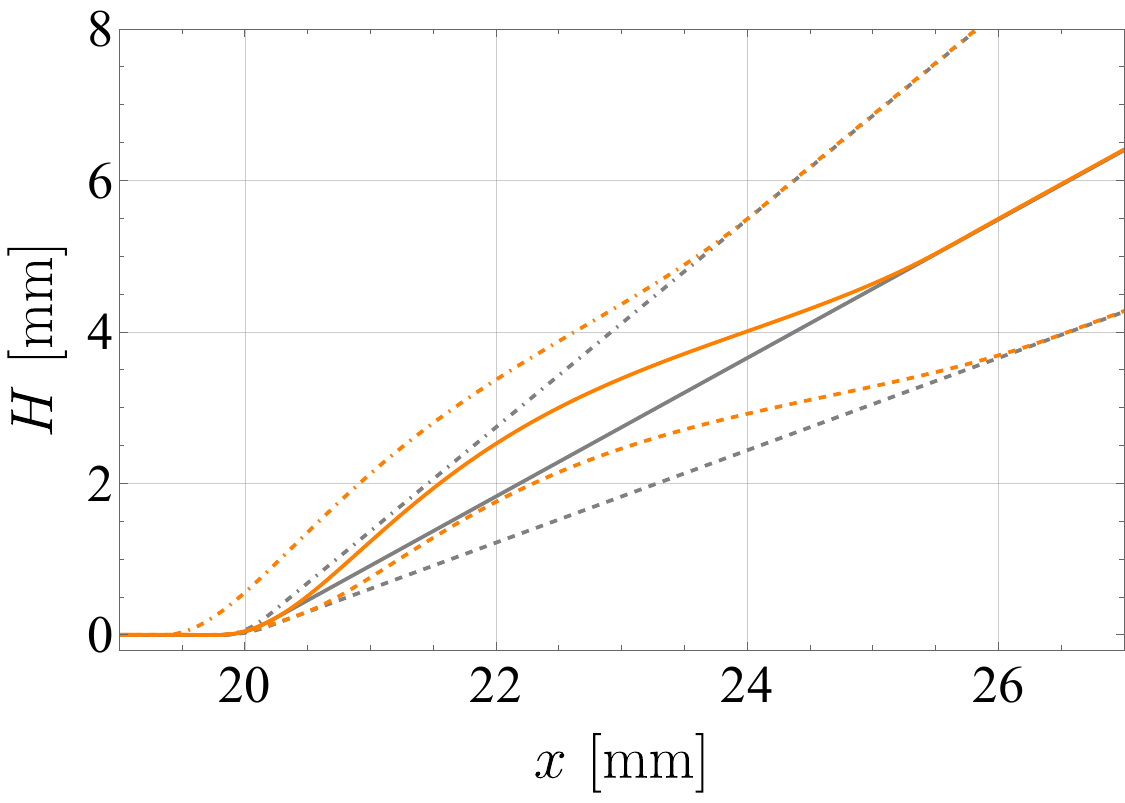}}
    \caption{(a) Climbing heights as a function of time for SAW amplitudes $A=1$~nm (black), $A=4$~nm (orange), and $A=6$~nm (green) for ramp slopes of $31.4^\circ$ (dashed lines), $42.4^\circ$ (solid lines), and $53.9^\circ$ (dot-dashed lines); colors shown here are as in Fig.~\ref{fig:climbingHeights_FinalTimes}(a) where a ramp of slope $42.4^\circ$ is used. (b) Final height profiles for the case of $A=4$~nm for each of the three different ramp slopes used in (a).}
    \label{fig:rampSlope}
\end{figure}

\begin{figure}
    \centering
    \subfigure[]{
    \includegraphics[width=0.48\linewidth]{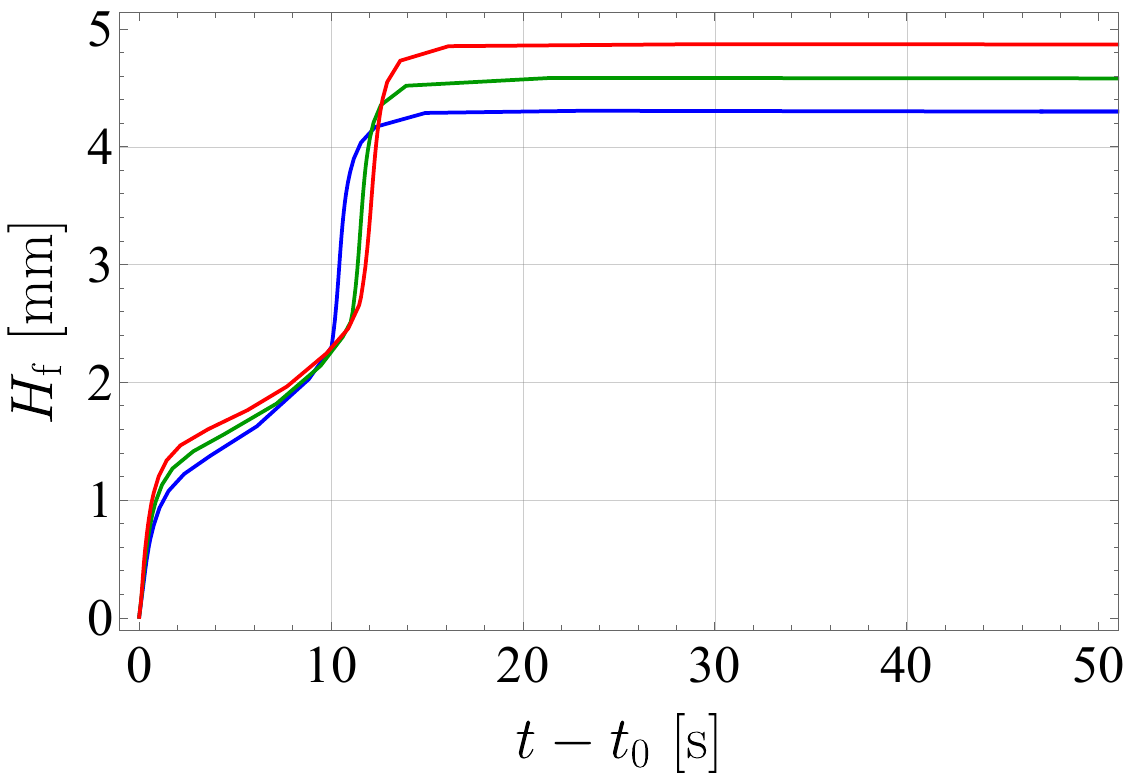}}
    \subfigure[]{
    \includegraphics[width=0.48\linewidth]{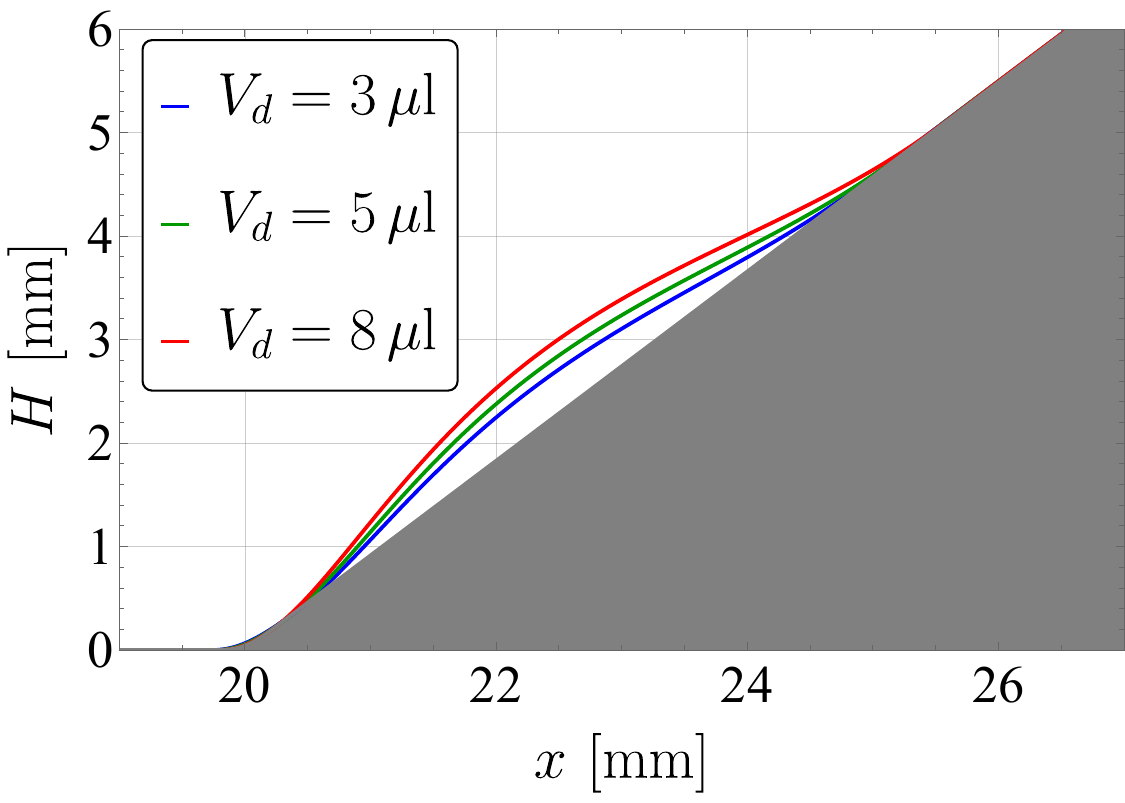}}
    \caption{(a) Climbing heights, $H_f$, of the drop's front as a function of time on the ramp obstacle for amplitude $A=4$ nm calculated for four different drop volumes, $V_d$: $3~\mu$l ({blue}), $5~\mu$l ({green}), and $8~\mu$l ({red}). (b) Steady-state thickness profile for each drop volume, $V_d$, for large times (i.e., when there is no flow).
    }
    \label{fig:rampVol}
\end{figure} 

\subsection{Spreading over a bump}  \label{sec:bump}
We next examine the spreading of an oil film over a bump-shaped obstacle, modeled after the experimental setup shown in Fig. \ref{fig:bump}. The bump is defined by the parabolic shape profile:
\begin{equation}
        \label{eq:sbump}
   s(x)= \left\{ 
        \begin{array}{ll}
             0 ,&  x <  x_{\rm o}-\frac{w_{\rm o}}{2} ,\\
            h_{\rm o}\frac{\sqrt{1-\Big(\frac{x-x_{\rm o}}{w_{\rm o}}+\frac{1}{2}\Big)^2}-\frac{\sqrt{3}}{2}}{1-\frac{\sqrt{3}}{2}} ,& 
            x_{\rm o}-\frac{w_{\rm o}}{2}\leq x \leq  x_{\rm o}+\frac{w_{\rm o}}{2},\\ 
             0 ,& x >   x_{\rm o}+\frac{w_{\rm o}}{2}, 
        \end{array} 
        \right.
\end{equation}
where, motivated by the experimental values,  we set $x_{\rm o}=25$, $h_{\rm o}=0.57$, and $w_{\rm o}=2.8$.

\begin{figure}[ht]
    \centering
    \includegraphics[height=0.22\linewidth]{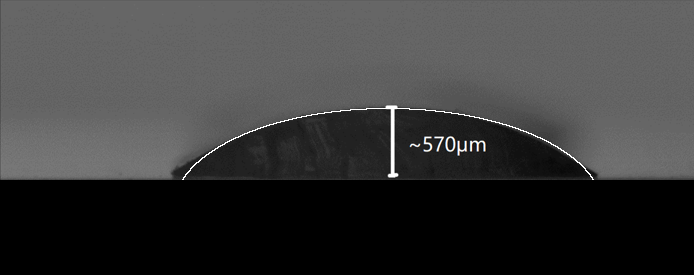}
    \caption{Side view of the bump obstacle used in experiments, with the circular arc employed in simulations overlaid to illustrate the close agreement between the experimental geometry and the model representation.}
    \label{fig:bump}
\end{figure}

\begin{figure}[t]
\centering
\subfigure[~$t=1$ s]{
\includegraphics[width=0.48\linewidth]{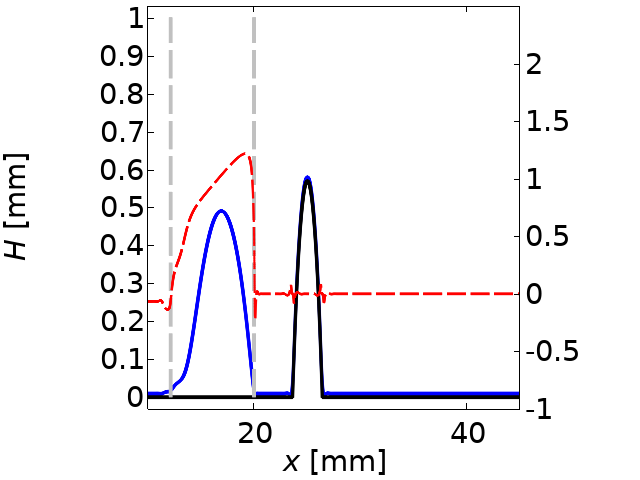}}
\subfigure[~$t=5$ s]{
\includegraphics[width=0.48\linewidth]{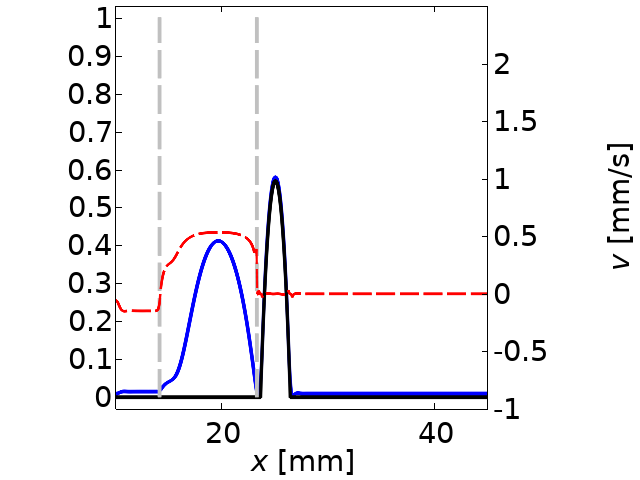}}
\subfigure[~$t=10$ s]{
\includegraphics[width=0.48\linewidth]{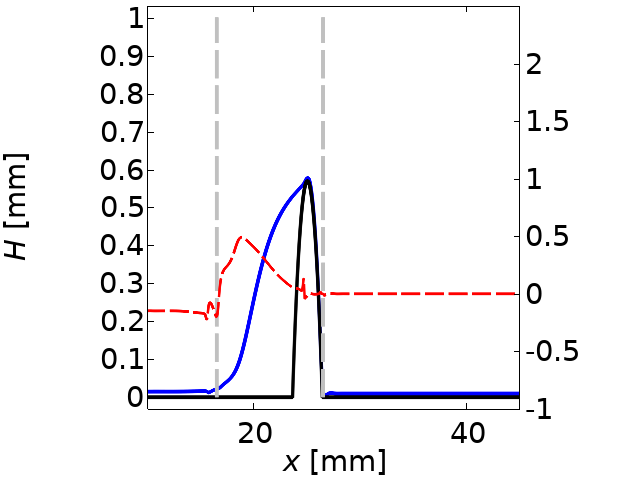}}
\subfigure[~$t=20$ s]{
\includegraphics[width=0.48\linewidth]{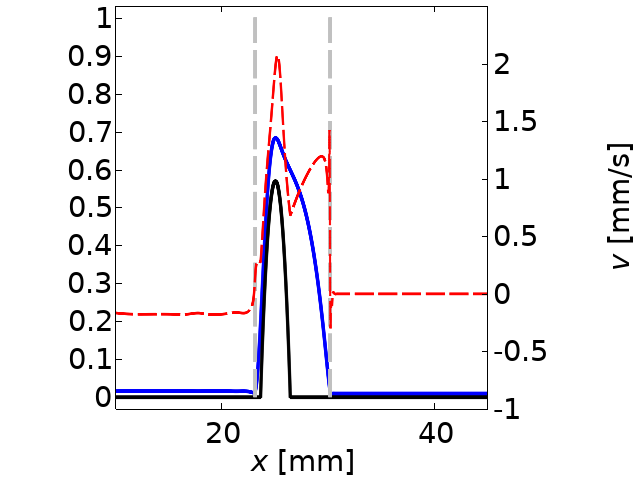}}
\subfigure[~$t=35$ s]{
\includegraphics[width=0.48\linewidth]{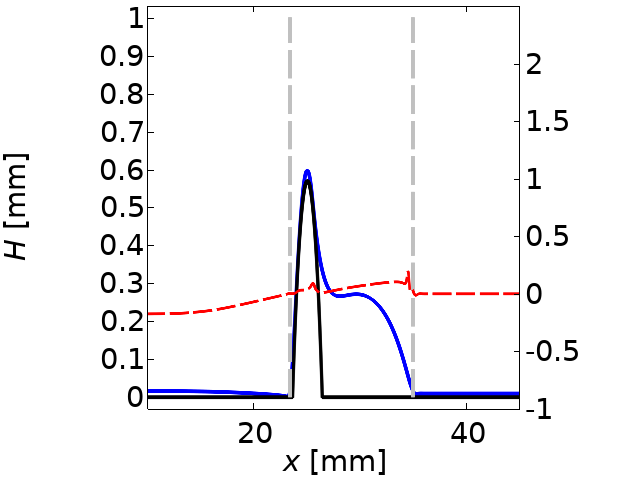}}
\subfigure[~$t=250$ s]{
\includegraphics[width=0.48\linewidth]{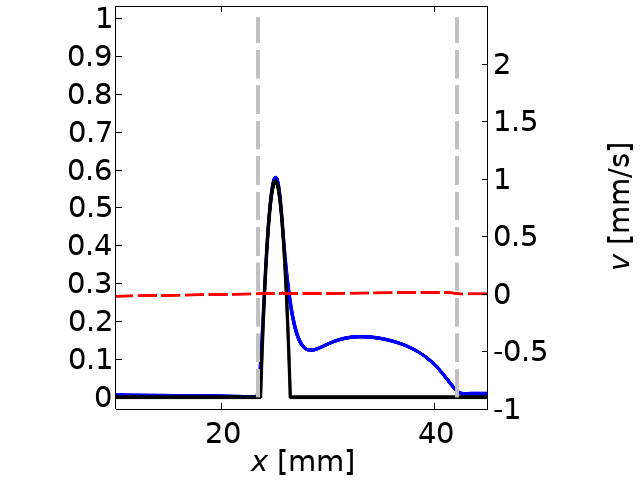}}
\caption{(a)-(f) Evolution of thickness profile ({blue} line) and thickness-averaged horizontal velocity component ({red}, dashed line) for $A=4$ nm at indicated times as the oil climbs a bump obstacle (black). The {gray} dashed lines denote the positions of $x_{\rm r}^\ast(t),x_{\rm f}^\ast(t)$. Full simulation videos are available upon request for a range of $A$ values.} 
\label{fig:bump_Snapshots}
\end{figure}

Figure~\ref{fig:bump_Snapshots} shows the evolution of the film profile and the thickness-averaged horizontal velocity component for $A=4$~nm. The gray dashed lines indicate the positions of $x_{\rm r}^\ast(t)$ and $ x_{\rm f}^\ast(t)$, the rear and front cutoff line positions at which attenuation starts and ends. The initial dynamics is the same as for a flat substrate~\cite{Li2024}. Once the deformed film reaches the bump obstacle and begins to climb, the horizontal velocity at the front decreases, as shown in Fig.~\ref{fig:bump_Snapshots}(b-c). This decrease occurs because the front of the film encounters the upward slope of the bump, where the local gravitational resistance increases and the SAW forcing becomes attenuated due to the larger total film thickness ($H=h+s$). Meanwhile, the rear of the film remains on the flat substrate to the left of the bump, where the SAW-induced stress is still strong, generating a net internal flow that drives fluid forward. As a result, the rear moves faster than the front, compressing the film. (This is inevitable in our 2D simulations due to mass conservation, but not in the 3D experiments, where the film can expand laterally.) The reduced width means that less acoustic attenuation occurs across the film, strengthening the SAW-induced stress felt by the front of the film and allowing it to climb fully over the obstacle. Note that, because the PDMS obstacle is assumed to have the same acoustic properties as the silicone oil, even after the film has traversed the obstacle, the SAW begins attenuating at the leftmost point of the obstacle (near $x=x_{\rm o}-w_{\rm o}/2$). Thus, the SAW-induced stress on the film significantly decreases when it is to the right of the obstacle; its front speed is strongly diminished, and the film width continually increases primarily due to capillary forces (silicone oil is fully wetting on lithium niobate). The interested reader can inquire about animations for different values of $A$. 

Figure~\ref{fig:contactLines} shows the positions of $x_{\rm r}^\ast(t),\, x_{\rm f}^\ast(t)$ (marking the extent of the SAW attenuation zone), the film front $x_{\rm f}(t)$, and the film rear $x_{\rm r}$ (defined by the point where $h(x_r)=h^*$); the gray dashed horizontal lines indicate the starting and ending points of the bump obstacle. Note the near-constant speed at which the rear of the film translates before reaching the obstacle, a feature consistent with results for the dynamics of a silicone oil film on a flat substrate~\cite{Li2024}; we point out the collapse of $x_{\rm r}$ and $x_{\rm r}^\ast$ onto a single curve is due to the fact that the rear never climbs onto the bump obstacle as a result of the precursor thickness and large slope of the obstacle (see Fig. \ref{fig:bump_Snapshots}). One might expect that because the governing equation is parabolic (and theoretically propagates information instantaneously), the rear of the film should feel the effects of the obstacle the moment the front reaches it, yet this clearly does not happen: the rear continues to move with near-constant speed well after the front reaches the obstacle. 
This is explained by the fact that the rate at which a disturbance becomes dynamically significant is controlled by the fourth-order capillary diffusion operator, which operates on a much slower time scale. Linearizing the nondimensional PDE (Eq.~\eqref{eq:ndGovEqMain}) about a region of nearly uniform thickness $h_0$ shows that disturbances spread over a nondimensional distance $W$ on the characteristic nondimensional time scale $t_{\rm d}\sim W^4/h_0^3$. For our simulations, we obtain the width and height of the film during this time as $W\approx 9.5$ and $h_0\in[.017,.4]$. Using the largest thickness (which gives the fastest possible time), the estimate yields a dimensional (found by multiplying by our time scale, $T$) disturbance time of $t_{\rm d}\approx 15$ minutes, far longer than the duration of the climbing event. Thus, the rear cutoff line cannot respond to the presence of the obstacle while the front is climbing until the rear itself gets much closer to the obstacle. 

Additionally, we point out that the length of the flat part of $x_{\rm f}^\ast(t)$ (green curve in Fig.~\ref{fig:contactLines}) corresponds to the climbing time of the oil film over the obstacle, discussed in some detail in Section \ref{sec:discussion} below. Based on our definitions of the cutoff lines (points at which $h+s=h^\ast$), once the oil front reaches the rear of the obstacle, the location of $x_{\rm f}^\ast(t)$ jumps to the front of the obstacle (due to the additional total effective film thickness from the obstacle). It stays at this point until enough of the bulk oil has fully climbed the obstacle, then begins further translating across the domain. This period, defined as the climbing time, exactly corresponds to the time it takes for the front contact line, $x_{\rm f}$, to pass from one side of the obstacle to the other. Finally, we point out the relative speeds of the front contact line before and after reaching the bump. Immediately after passing over the bump, the front contact line starts to move with larger velocity due to the added gravitational contribution (which now aids movement of the front rather than hindering it during climbing). At long times, however, the front slows due to horizontal attenuation of the SAW increasing as the front moves further ahead of the the obstacle. At long times after fully climbing, the front moves almost entirely due to capillary spreading. 

\begin{figure}[ht!]
    \centering
    \includegraphics[width=0.5\linewidth]{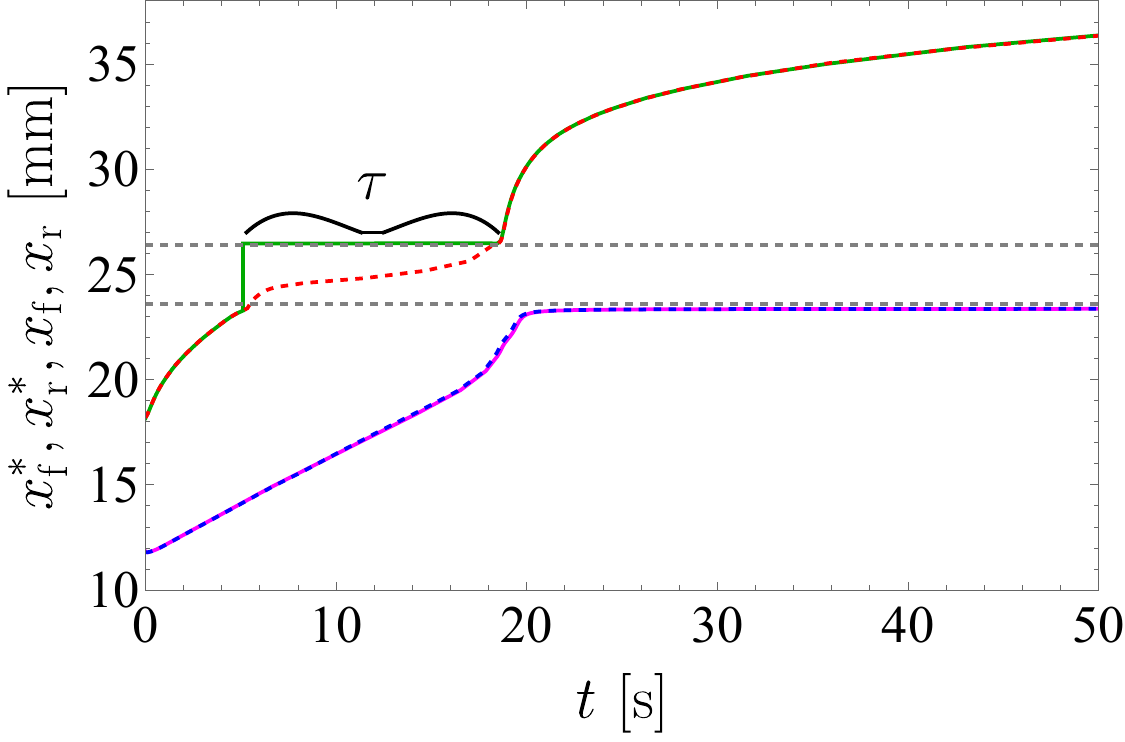}
    \caption{Positions of the front cutoff line $x_{\rm f}^\ast$ ({green}), rear cutoff line $x_{\rm r}^\ast$({magenta}), the front contact line $x_{\rm f}$ ({red}, dashed), and the rear contact line $x_{\rm r}$ (blue, dashed) of the oil film for the case of $A=4$ nm shown in Fig. \ref{fig:bump_Snapshots} for the bump obstacle. The dashed gray lines indicate the start and end positions of the bump obstacle. The climbing time of the oil film over the obstacle is denoted by $\tau$ as the horizontal section of the green curve.}
    \label{fig:contactLines}
\end{figure}

\section{Discussion}
\label{sec:discussion}
Here we discuss comparisons between experiment and theory, first for the ramp obstacle experiments and then the bump obstacle. Figure~\ref{fig:ramp_expSim} shows a comparison of the maximum climbing heights for experimental and simulation results over a ramp obstacle as a function of the SAW displacement amplitude ($A_n$ for experiments and $A$ for simulations).
Both experiments and simulations indicate a monotonic dependence of climbing height on SAW amplitude. Consistent with the findings of Fasano et al. for dynamics on a flat substrate~\cite{Li2024} (and as discussed in more detail in that work), the experimental trend is reproduced in our simulations for appreciably larger values of the SAW amplitude, i.e. $A>A_n$. Essentially, while the theory captures the qualitative trend in the experiment, it underpredicts the SAW-induced stress in the oil film. While it is not clear why this difference arises, some plausible reasons, in addition to possible differences between $A$ and $A_n$ \cite{Royer1996} include: (i) the potential relevance of inertial effects that are not included in the present model~\cite{ZAREMBO1971,Orosco2021UnravelingTC,Dubrovski_Friend_Manor_2023}; (ii) the omission of contributions from the boundary layer flow near the solid surface, i.e. the Rayleigh law of streaming; (iii) the capillary waves that appear at the free surface of the film \cite{LIGHTHILL:1978p12,Morozov17}; (iv)  the use of a two-dimensional model to understand a three-dimensional system; (v) the neglect of the ultrasonic wave reflection off the free surface of the liquid (i.e. acoustic radiation pressure).

\begin{figure}
    \centering
    \includegraphics[width=0.5\linewidth]{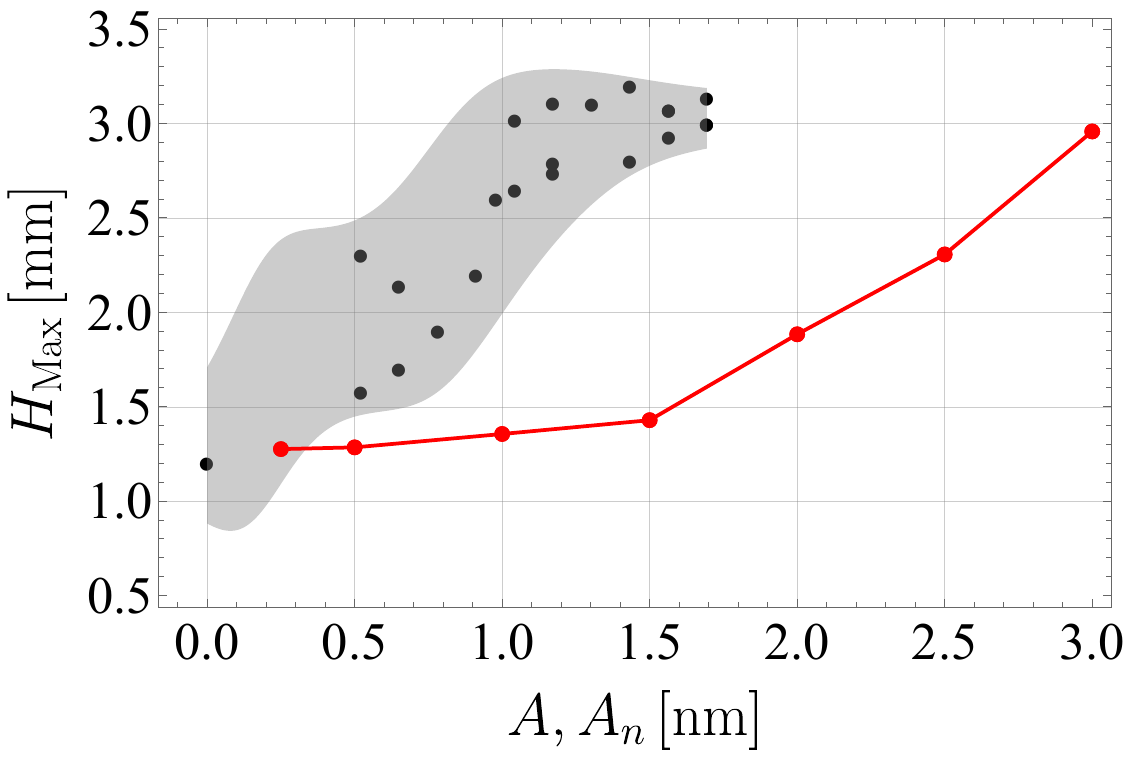}
    \caption{Maximum height on the ramp obstacle reached by the film front, $H_{\text{max}}$, as a function of the SAW amplitude, $A_n$ (experiments) or $A$ (simulations), as obtained from the experiments (black circles) and simulations (red points connected by a solid line). The shaded gray area indicates the general trend of the experimental data.}
    \label{fig:ramp_expSim}
\end{figure}

Moving now to the comparison with the bump obstacle experiments, Fig.~\ref {fig:climbingTime} shows the climbing times, $\tau$, for full traversal of four different height bumps, for both experiments (black) and simulations (red), on a log-log scale. For the simulations, $\tau$ is defined as the time taken for $x_{\rm f}$, the front contact line, to pass from one side of the obstacle to the other (i.e. the flat portion of the green curve in Fig.~\ref{fig:contactLines}). The dashed lines in Fig.~\ref{fig:climbingTime} show power-law fits for both experimental and simulation data. As is evident, the climbing time decreases with increasing SAW amplitude for all bump sizes. The model captures this trend qualitatively well for the bump shapes considered, in particular the power law dependence $\tau=aA^b$ (see the values of $b$ obtained in the fitted lines in Table \ref{tab:exponents}). The values of $a$ in the fits show some discrepancy between theory and experiment, in large part due to the appreciably larger SAW amplitude required in simulations to produce similar dynamics already noted. 

\begin{figure}[!h]
    \centering
    \subfigure[$h_{\rm o}=0.57, w_{\rm o}=2.8$]{\includegraphics[width=0.48\linewidth]{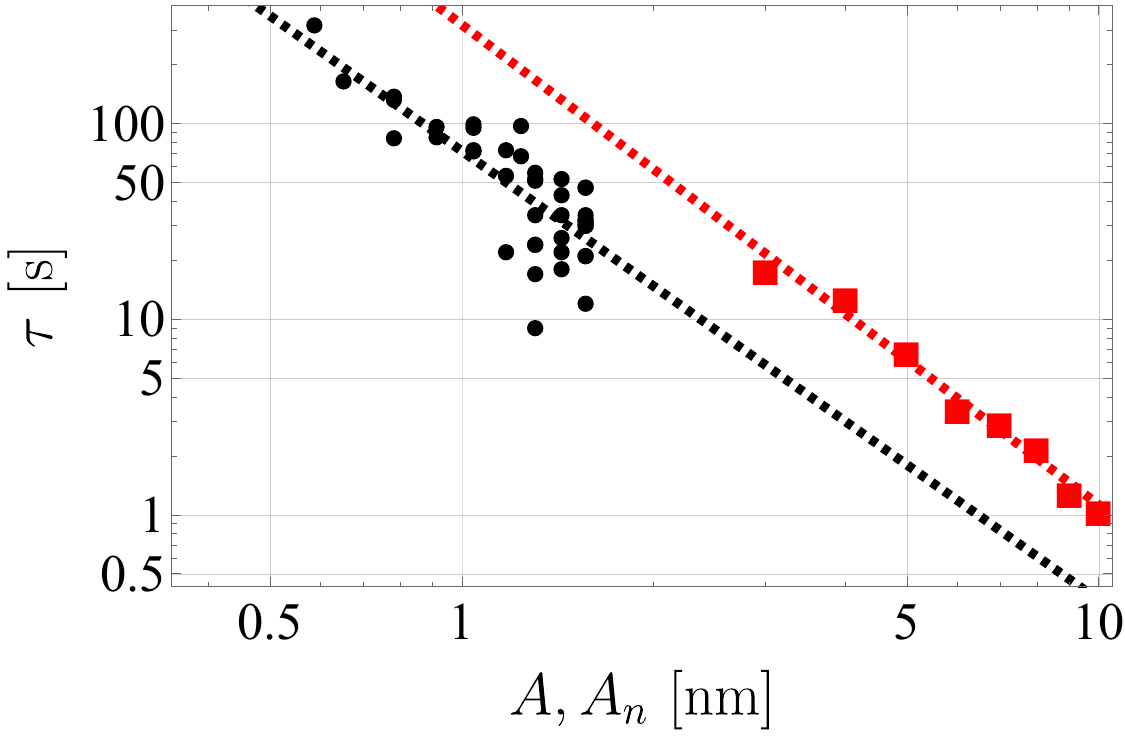}}
    \subfigure[ $h_{\rm o}=0.26, w_{\rm o}=1.8$]{\includegraphics[width=0.48\linewidth]{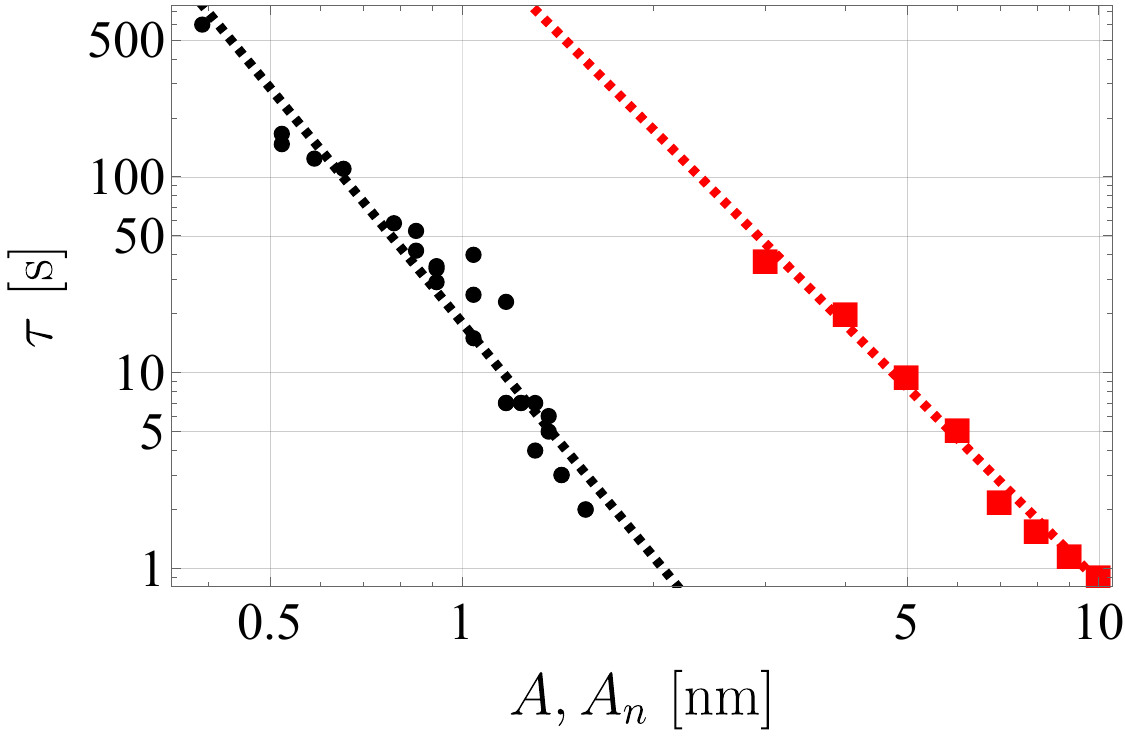}}
    \subfigure[$h_{\rm o}=0.71, w_{\rm o}=2.5$]{\includegraphics[width=0.48\linewidth]{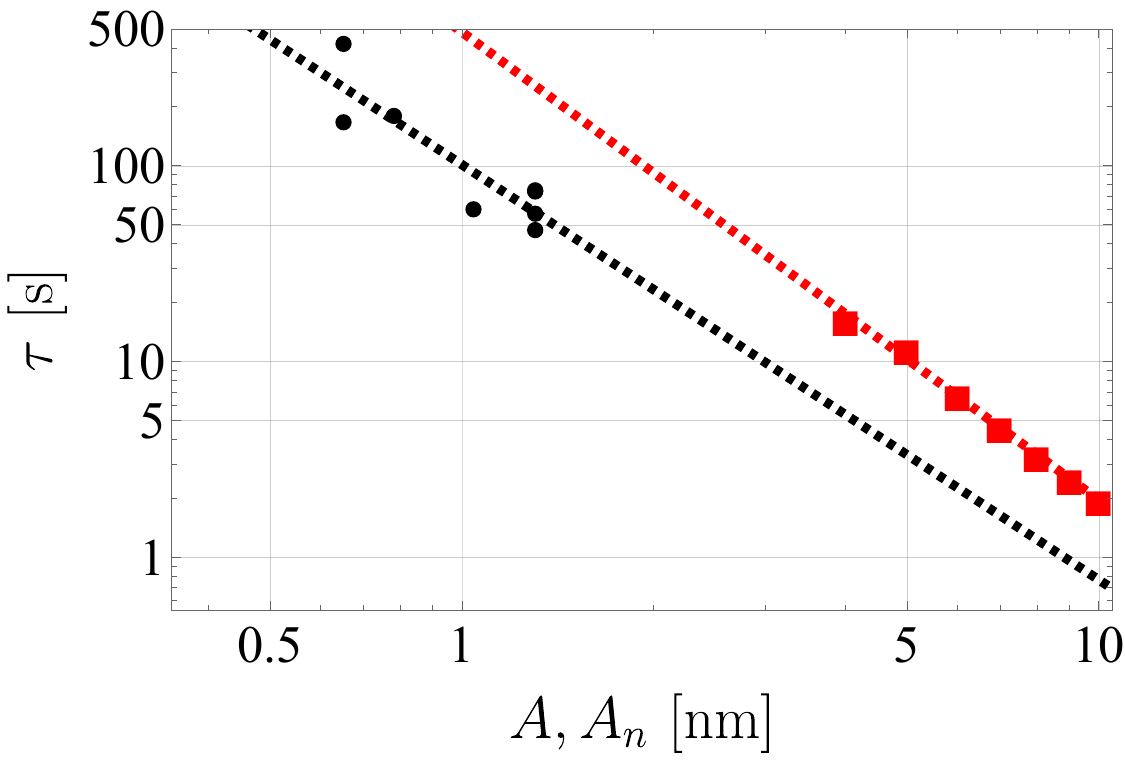}}
    \subfigure[$h_{\rm o}=0.65, w_{\rm o}=2.5$]{\includegraphics[width=0.48\linewidth]{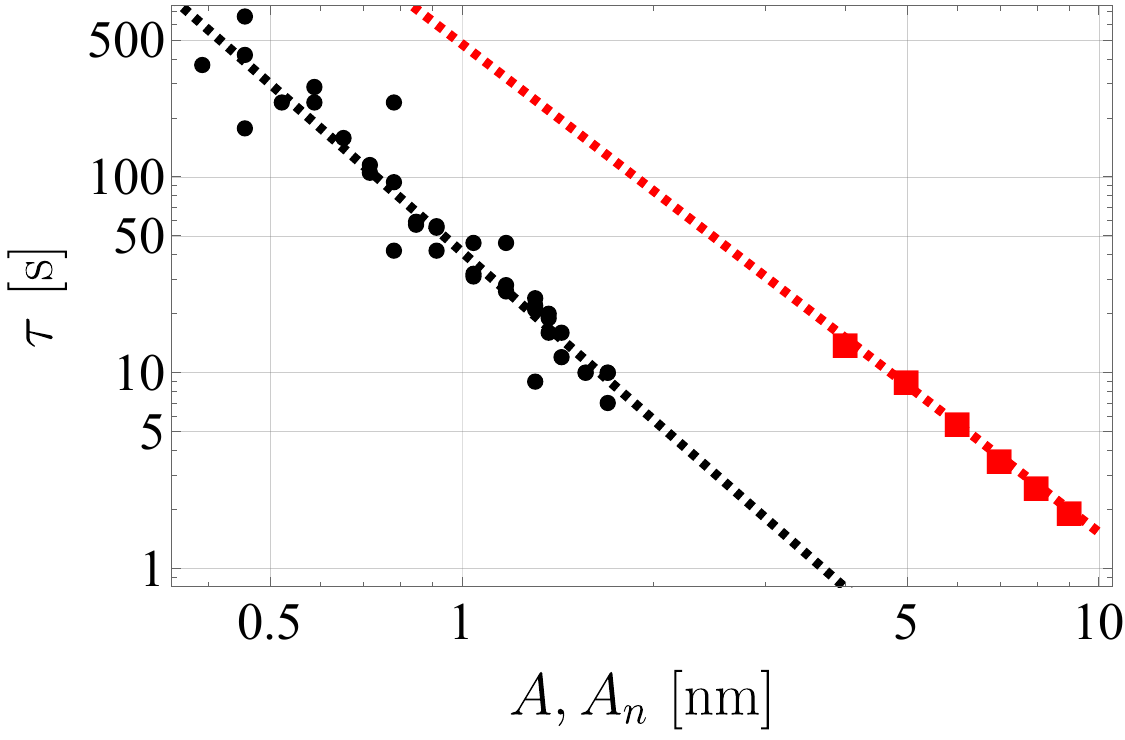}}
    \caption{Climbing time, $\tau$, from the experiments (black dots) and simulations (red squares) for selected values of the SAW amplitude, $A_n$ (experiments) or $A$ (simulations), for four different sized bump obstacles. The black (red) dashed line corresponds to an approximating function for the experimental (numerical) data of the type, $\tau=a\bar A^b$, where $\bar A$ stands for either $A$ or $A_n$. The exponents $b$ for both experimental and numerical results are presented in Table \ref{tab:exponents}.}
    \label{fig:climbingTime}
\end{figure}

\begin{table}[!h]
  \centering
  \caption{Power-law exponents from experiments, $b_{\rm exp}$, and simulations, $b_{\rm sim}$, for the fitting lines, $\tau = a_{\rm sim} A^{b_{\rm sim}}$, $\tau = a_{\rm exp} A_n^{b_{\rm exp}}$, corresponding to the four panels of Fig.~\ref{fig:climbingTime}. The relative errors between experiments and simulations are $7-17$\%.}
  \label{tab:exponents}
  \begin{tabular}{@{}ll
                  S[table-format = -1.2]
                  S[table-format = -1.2]@{}}
    \toprule
    Fig.~\ref{fig:climbingTime} Panel & Obstacle $(h_o,\,w_o)$ & {$b_{\text{exp}}$} & {$b_{\text{sim}}$} \\
    \midrule
    (a) & $(0.57,\,2.8)$ & -2.29 & -2.45 \\
    (b) & $(0.26,\,1.8)$ & -3.98 & -3.30 \\
    (c) & $(0.71,\,2.5)$ & -2.12 & -2.39 \\
    (d) & $(0.65,\,2.5)$ & -2.85 & -2.49 \\
    \bottomrule
  \end{tabular}
\end{table}

\section{Conclusions} \label{sec:conclusions}

We have presented experimental and theoretical studies of the SAW-driven motion of a thin, viscous silicone oil film over solid obstacles. The experiments reported here are, to our knowledge, the first to show that surface acoustic waves can drive a millimeter-scale film to climb over well-defined topographical features. To interpret and predict this behavior, we developed a new long-wave-type model that incorporates substrate geometry into both the capillary pressure and the SAW-induced stress in a self-consistent manner. The long-wave model derived in Appendix \ref{app:lub} extends prior SAW thin-film models by allowing the acoustic attenuation to depend on the evolving local film thickness above obstacles, and accounts for topographical contributions to the free-surface curvature, to the gravitational terms, and to the acoustic attenuation.

The simulations reveal a physically intuitive steady state when the film climbs a ramp: the film reaches a finite equilibrium climbing height at which capillarity, gravity, and SAW-induced stress are balanced. This balance in simulations is due to our treatment of the obstacle (cured PDMS) as having the same acoustic properties as the silicone oil phase itself; if the attenuation length of the SAW within the obstacle is much larger than that of silicone oil, then one would expect the fluid to climb much higher. Moreover, the model predicts that increasing the ramp's slope increases climbing height, despite the greater gravitational opposition. This counterintuitive result arises because a steeper ramp narrows the film footprint, reducing the horizontal extent over which the SAW attenuates and therefore strengthening the forward acoustic driving. 

The transient dynamics also displays a clear separation of roles between the front and rear of the film. The rear contact line advances at nearly constant speed for a significant portion of the climb, even after the front of the film has begun to deform or slow upon encountering the obstacle. This occurs because the rear portion of the film continues to experience the undisturbed SAW field until it physically interacts with the obstacle, resulting in decoupling of the front and rear film motion. This can be understood by examining the time scale on which disturbances travel through the fluid.

For the bump obstacles employed, the developed model qualitatively captured the power-law dependence of climbing time, $\tau$, on the SAW amplitude, with fairly good accuracy across a range of obstacle parameters (see Table \ref{tab:exponents}). This indicates that, despite the significant simplifications made in deriving the governing equations, our model retains the most important physical mechanisms. 

Taken together, these results provide a predictive framework for SAW-driven thin-film motion over structured substrates, grounded in both experiment and theory. They identify the roles of geometry, SAW-induced stress, and capillarity in determining whether and how a viscous film climbs over an obstacle. This framework lays the foundation for future studies of SAW-driven spreading in multilayer films, films containing internal interfaces, and patterned substrates where more complex interfacial dynamics are expected. Such setups provide a promising testbed for using SAW to drive coating flows over objects, offering a new paradigm for the coating industry. \\

\noindent\vspace{0.2in}
{\bf Acknowledgments}

This work was supported by the donors of ACS Petroleum Research Fund under PRF\# 62062-ND9 and by BSF grant No. 2020174. J.A.D acknowledges support from Consejo Nacional de Investigaciones Científicas y Técnicas (CONICET, Argentina) with Grant PIP 02114-CO/2021 and Agencia Nacional de Promoción Científica y Tecnológica (ANPCyT, Argentina) with Grant PICT 02119/2020.

\appendix
\renewcommand{\theequation}{\thesection.\arabic{equation}}

\section{Theoretical model for flat substrate} \label{app:SpreadingModel}
This appendix summarizes the theoretical framework for SAW-driven flow on a flat substrate, which forms the starting point for the obstacle modified long-wave model developed in Appendix \ref{app:lub} and presented in Section \ref{sec:model}. The full derivation, including all intermediate steps and explicit expressions for the acoustic fields, is given in \cite{Li2024} following the approach of \cite{vanneste_streaming_2011} and \cite{campbell_propagation_1970}. Here, we retain only the essential elements required to identify the form of the second-order acoustic forcing and clarify how substrate topography alters the long-wave reduction. Throughout this appendix, the substrate is assumed flat; the role of topographical features enters only in Appendix \ref{app:lub} through modified boundary conditions.

We consider a two-dimensional Newtonian fluid layer occupying a domain in the $(x,z)$--plane, traversed by a sound wave generated by a Rayleigh-type-SAW traveling in a neighboring solid. The problem is governed by the Navier-Stokes and continuity equations, written in the fluid domain. The acoustic Mach Number $M_a \equiv\omega A/c$ ($c$ is the phase velocity of the acoustic field in the fluid) is small, allowing the velocity, pressure, and density to be expanded asymptotically in powers of $M_a$. The leading order state corresponds to a quiescent fluid ($\mathbf{v}_0 =\mathbf{0}$) of uniform density $\rho_0$, while the first and second order problems describe oscillatory acoustic motion and steady streaming, respectively. Gravity is assumed to enter at second order in $M_a$, consistent with the separation of time scales between acoustic oscillations and slow evolution of the free surface.

\subsection{First order solution}
\label{sec:first}

At first order, the flow corresponds to the oscillatory acoustic field induced in the liquid by the propagating leaky SAW. Because the flow at this order is dominated by high-frequency oscillations and inertia, viscous effects are weak except within a thin boundary layer adjacent to the solid surface. For the film thicknesses and frequencies relevant to the present experiments, this viscous boundary layer is much thinner than the bulk of the liquid. As a result, vorticity generated near the substrate remains localized and does not significantly influence the bulk motion. Motivated by this scale separation, the first order velocity field is assumed to be irrotational in the bulk of the liquid, and is therefore represented as $\mathbf{v}_1=\nabla\phi$ where $\phi$ is the acoustic velocity potential. This assumption is standard in treatments of bulk acoustic streaming and is consistent with prior theoretical and experimental studies of leaky SAW-driven flows. 

Substituting this form into the first order governing equations yields a damped wave equation for $\phi$, accounting for viscous attenuation of the acoustic field within the fluid. The acoustic excitation enters through a boundary condition at the solid-liquid interface, which prescribes the normal velocity induced by the Rayleigh-type SAW propagating along the substrate. This boundary condition is obtained from classical elastic wave theory in solids, as described by Royer \& Dieulesaint \cite{Royer1996}, with the modification that the wave is leaky due to coupling with the fluid.

Solving the resulting boundary-value problem leads to a time-harmonic acoustic potential that propagates in the direction of the SAW and decays exponentially both normal to the substrate and along the direction of propagation. The associated first order velocity and pressure fields oscillate at the SAW frequency and have zero mean when averaged over one acoustic period. The acoustic velocity potential is given by 
\begin{equation}
\label{eq:HelmholtzEqSol}
\phi=\frac{- A\omega}{\sqrt{\kappa_{\rm s}^2-\kappa^2}}\exp({i\omega t})\exp{\left(-i\kappa_{\rm s} x\right)}
\exp{\left(-z\sqrt{\kappa_{\rm s}^2-\kappa^2} \right)},
%\label{eq:phi_vel}
\end{equation}
where $\kappa$ and $\kappa_s$ are complex wavenumbers of the SAW in the liquid and solid respectively. For more details on their meaning and physical value, we point the interested reader to \cite{Li2024}. Additionally, the first order fields are identical to those derived in \cite{Li2024} and are not reproduced here for brevity and are not directly modified by the presence of substrate topography. The sole role of the first order solution in the present work is to generate, through quadratic nonlinearities, a non-zero time-averaged Reynolds stress that appears as a body force in the second order problem discussed next.

\subsection{Second order solution}
\label{sec:second}

At second order ($O(M_a^2)$), time-averaging over one acoustic period yields a steady Stokes flow $\mathbf{v}_2$ driven by the divergence of the Reynolds stress associated with the first order acoustic field. The resulting governing equations are
\begin{eqnarray}
\label{eq:NSv}
&-\nabla \left<p_2\right> + {\mathbf{F}}_s -\rho_0 g\mathbf{e}_z + \mu \nabla^2 \left<\mathbf{v}_2\right>=0, \\
\label{eq:cont_v2}
&\nabla\cdot\left<{\bf v}_2\right>=0.
\end{eqnarray}
where $\mathbf{F}_s$ is the SAW-induced body force. The acoustic forcing takes the form of exponentially attenuated body forcing in the horizontal and vertical directions, 
\begin{eqnarray}
    F_{s,x} &=& C_x P_0 \, e^{-2 (k_{\rm s,i} x + K_z z)}, 
\label{eq:Fsx}\\
    F_{s,z} &=& C_z P_0 \, e^{-2 (k_{\rm s,i} x + K_z z)},
\label{eq:Fsz}
\end{eqnarray}
where $P_0=\rho_0A^2\omega^2$ sets the characteristic stress scale. The coefficients $C_x,C_z$ and attenuation factors $k_{\rm s,i},K_z$ depend on the acoustic properties of the fluid and substrate. These expressions are identical to those derived in \cite{Li2024}. Their role in the present paper is to define the functional dependence of the driving stress that enters the long-wave-type model.

\section{Long-wave model for obstacle problem} 
\label{app:lub}
In this section, we provide a step-by-step description of deriving Eq.~\eqref{eq:ndGovEqMain} from the second-order system defined by Eqs.~\eqref{eq:NSv},~\eqref{eq:cont_v2},~\eqref{eq:Fsx}, and~\eqref{eq:Fsz}. In what follows, we focus on a simplified two-dimensional geometry that assumes translational invariance in the transverse $y$-direction (i.e. that $\partial/\partial y=0$); the discussion of three-dimensional effects will be presented elsewhere. As previously mentioned, the obstacle enters the derivation at this stage through the various boundary conditions used to translate this Stokes flow system of equations to a single PDE for the height of the oil film, $h(x,t)$.

The resulting nondimensional system, using the scales presented in Sec.~\ref{sec:nd}, in component form is (after letting $\langle \mathbf{v}_2\rangle=\mathbf{v}=(v_x,v_z)^T$ and $\langle p_2\rangle=p$)
\begin{eqnarray}
0&=& -\frac{\partial p}{\partial x} + \frac{1}{3} \frac{\partial^2 v_x}{\partial z^2} + \mathcal{S}C_x \, \psi(x,z),
\label{eq:ndNSx}\\
0 &=& -\frac{\partial p}{\partial z} - \text{Bo}+\mathcal{S}C_z\, \psi(x,z).
\label{eq:ndNSz}
\end{eqnarray}
Note that we have used a standard long-wave model to rewrite the viscous term as the second derivative of the horizontal component of the velocity with respect to $z$ since it is the dominant term in the context of thin-films considered here \cite{Kondic2003}. The nondimensional parameters and the nondimensional function, $\psi$, in the problem are defined as
\begin{equation}
    \text{Bo} = \frac{\rho_0gL^2}{\gamma}, \quad \mathcal{S} = \frac{P_0L}{\gamma}, \quad \psi(x,z) = e^{-2(k_{\rm s,i}x+K_zz)}.
\end{equation}
To solve for the pressure, we need to specify a boundary condition at the free surface, which we do using the following nondimensional Laplace pressure boundary condition ~\cite{stillwagon1988fundamentals}:
\begin{equation}
p\bigg\rvert_{z=h+s} = -\kappa_1 = - s''(x)-h''(x),
\label{eq:psup}
\end{equation}
Note that we simplify the curvature of the free surface in the boundary condition in the spirit of the long-wave model. One can then solve for the pressure explicitly by integrating Eq. \eqref{eq:ndNSz} with respect to $z$ and imposing Eq. \eqref{eq:psup}. Then, differentiating the expression for pressure with respect to $x$ and substituting it into Eq. \eqref{eq:ndNSx} one is left with a second-order PDE for $v_x$, 
\begin{eqnarray}
    \frac{\partial^2 v_x}{\partial z^2} &=& 3\bigg(- \kappa _1'+ \text{Bo}\, (h'+s') +\frac{\mathcal{S}C_z}{2K_z}\frac{\partial}{\partial x}\Big(\psi(x,h+s)\Big)\bigg) - \mathcal{C} \frac{3\mathcal{S}}{K_z}\psi(x,z)
    \label{eq:dv2}
\end{eqnarray}
Imposing zero shear stress at the free surface ($\partial v_x/\partial z\rvert_{z=h+s}=0$) and no-slip at the obstacle-oil interface ($v_x\rvert_{z=s}=0$), we can integrate Eq.~\eqref{eq:dv2} twice with respect to $z$ to find an explicit expression for $v_x$. As is standard in long-wave models, we next need to average over the thickness of the film to remove the $z$-dependence of $v_x$ by defining $v=(1/h)\int_s^{h+s}\, v_x\, dz$. Finally, with this expression of $v$, the mass conservation equation, namely $h_t+(vh)_x=0$, can be solved numerically. It is convenient to rewrite this equation in a slightly modified form by defining an effective pressure,
\begin{equation}
    \mathcal{P} = -\kappa_1+ \text{Bo}\, (h+s)+\frac{\mathcal{S}C_z}{2K_z}\psi(x,h+s). \label{eq:ndEffP}
\end{equation}
so that the governing equation for the oil film is
\begin{eqnarray}
    \pder[h]{t} + \pder{x}\bigg[-h^3 \pder[\mathcal{P}]{x}
    -\mathcal{C}\frac{3\mathcal{S}}{8K_z^4}\psi(x,h+s)\Big(2K_z^2h^2-1+e^{2K_zh}(1-2K_zh)\Big)\bigg] = 0 \label{eq:ndGovEq}.
\end{eqnarray}
It is easily verifiable that in the case of negligible gravity and no SAW-induced stress, this equation reduces to the classic case studied by Stillwagon and Larson~\cite{stillwagon1988fundamentals}. Further, in the case of only gravity and capillarity considered, we have agreement with Park and Kumar~\cite{park2017droplet} (where one must neglect disjoining pressure and consider a flat substrate in their equations).

\bibliographystyle{RS}
\bibliography{films.bib}

\end{document}